\documentclass{jpp}
\usepackage{graphicx}
\usepackage{epstopdf, epsfig}
\usepackage{hyperref}
\usepackage[utf8]{inputenc}
\usepackage{bm}
\usepackage{amsmath}
\usepackage{xcolor}
\newcommand{\partder}[2]{\frac{\partial #1}{\partial #2}}
\newcommand{\der}[2]{\frac{d #1}{d #2}}

\usepackage{cleveref}
\crefname{equation}{}{}
\crefname{figure}{figure}{figures}
\crefname{appendix}{Appendix}{Appendices}
\crefrangelabelformat{equation}
{(#3#1#4--#5{#2}#6)}
\usepackage{subcaption}

\title{Adjoint approach to calculating shape gradients for 3D magnetic confinement equilibria}
\author{Thomas Antonsen, Jr.\aff{1}\corresp{antonsen@umd.edu}, Elizabeth J. Paul\aff{1}, Matt Landreman\aff{1}}
\affiliation{\aff{1} Institute for Research in Electronics and the Applied Physics, University of Maryland, College Park, MD 20742, USA}

\begin{document}

\maketitle

\begin{abstract}
    The shape gradient quantifies the change in some figure of merit resulting from differential perturbations to a shape. 
    Shape gradients can be applied to gradient-based optimization, sensitivity analysis, and tolerance calculation. An efficient method for computing the shape gradient for toroidal 3D MHD equilibria is presented. The method is based on the self-adjoint property of the equations for driven perturbations of MHD equilibria and is similar to the Onsager symmetry of transport coefficients. Two versions of the shape gradient are considered. One describes the change in a  figure of merit due to an arbitrary displacement of the outer flux surface; the other describes the change in the figure of merit due to the displacement of a coil. The method is implemented for several example figures of merit and compared with direct calculation of the shape gradient. In these examples the adjoint method reduces the number of equilibrium computations by factors of $\mathcal{O}(N)$, where $N$ is the number of parameters used to describe the outer flux surface or coil shapes. 
\end{abstract}

\section{Introduction}


The design of stellarator magnetohydrodynamic (MHD) equilibria requires optimizing within a high dimensional space due to the fully 3-dimensional nature of the magnetic field configuration and the sensitive dependence of charged particle trajectories in such field configurations \citep{Boozer2015}. While there are many possible choices for the space in which to optimize, a common choice is the space of the shape of the outer boundary of the plasma \citep{Hirshman1999,Drevlak2018}. The confining electromagnetic coils must then be designed to reproduce the desired plasma boundary. This approach was used to design Wendelstein 7-X \citep{Grieger1992} and the Helically Symmetric Experiment (HSX)  \citep{Anderson1995}. An alternative approach is to optimize the shape of electromagnetic coils directly to minimize an objective function which includes functions of the equilibria \citep{HansonCary}. This can be performed with the merged STELLOPT/COILOPT code, which was used in the late stages of the NCSX design \citep{Strickler2004}. 
To navigate such spaces, it is often useful to employ gradient-based optimization techniques. For this reason, it is desirable to compute derivatives with respect to shape. 

As the target optimized configuration can never be realized exactly, an analysis of the sensitivity to perturbations, such as errors in coil fabrication or assembly, is central to the success of a stellarator. Tight tolerances have proven to be a significant driver of the cost of stellarator experiments \citep{Strykowsky2009,Klinger2013}; thus an improvement to the algorithms used to conduct sensitivity studies can have great impact on the field. To quantify the coil tolerances for flux surface quality of LHD \citep{Yamazaki} and NCSX \citep{Brooks2003,Williamson2005}, perturbations of several distributions were manually applied to each coil. Sensitivity analysis can also be performed with analytic derivatives. Numerical derivatives with respect to tilt angle and coil translation of the CNT coils have been used to compute the sensitivity of the rotational transform on axis \citep{Hammond2016}. Analytic derivatives have recently been applied to study coil sensitivities of the CNT stellarator by considering the eigenvectors of the Hessian matrix \citep{Zhu2018}. Thus, in addition to gradient-based optimization, derivatives with respect to shape can be applied to sensitivity analysis.

The gradients of figures of merit with respect to shape have often been represented as derivatives with respect to whichever quantities parameterize the shape. Examples of such quantities are the amplitudes ($R_{mn}^c$, $Z_{mn}^s$)  in the double Fourier series for the cylindrical coordinates of a toroidal surface. Another way to represent  derivatives with respect to shape is the shape gradient \citep{Landreman2018}, which provides a local and coordinate-independent form. Consider any scalar figure of merit, $f$, that depends on a 3D MHD equilibrium solution. We can consider $f$ to be a functional of the shape of the outer boundary of the plasma, $S_P$. In this case, a differential change to the boundary, $\delta \textbf{r}$, causes a corresponding change to the figure of merit, $\delta f$,
\begin{gather}
    \delta f(S_P;\delta \textbf{r}) = \int_{S_P} d^2 x \, S \delta \textbf{r} \cdot \textbf{n}.
    \label{eq:shape_gradient_surface}
\end{gather}
Here $\textbf{n}$ is the outward unit normal and $S$ is the shape gradient. The shape gradient quantifies the local linear sensitivity of a figure of merit to differential perturbations of the shape. As tangential displacements to $S_P$ do not cause any changes to $f$, $\delta f$ only depends on the normal component of $\delta \textbf{r}$. The Hadamard-Zol\'{e}sio structure theorem \citep{Delfour2011} states that under certain assumptions of smoothness, the shape derivative of a functional can be written in the form of  \eqref{eq:shape_gradient_surface}. This can be thought of an instance of the Riesz representation theorem, which states that any linear functional, such as $\delta f(\delta \textbf{r})$, can be written as an inner product over the appropriate space \citep{Rudin2006}. The motivation for the form of \eqref{eq:shape_gradient_surface} is discussed in more detail in section 2 of  \cite{Landreman2018}. 

If we consider $f$ to be a functional of the shape of the electromagnetic coils, a differential change to the coils, $\delta \textbf{r}_C$, will cause a corresponding change to the figure of merit, $\delta f$,
\begin{gather}
    \delta f(C;\delta \textbf{r}_C) = \sum_{k} \int_{C_k} dl \, \textbf{S}_k \cdot \delta \textbf{r}_{C_k}.
    \label{eq:shape_gradient_coil}
\end{gather}
Here the sum is taken over the coils, $C_k$ is a curve describing the filamentary shape of each coil, $C = \{ C_k \}$, and $\delta \textbf{r}_C = \{ \delta \textbf{r}_{C_k} \}$. The shape gradient for coil $k$ is $\textbf{S}_k$. As tangential displacements to the coil do not change $f$, $\textbf{S}_k$ must be perpendicular to the tangent vector along $C_k$.

As the shape gradient provides local sensitivity information, it can be used to quantify engineering tolerances with respect to the displacement of coils or magnetic perturbations. The shape gradient representation can be computed from parameter derivatives by solving a small linear system \citep{Landreman2018}. 

However, computing parameter derivatives can often be computationally expensive, as numerical derivatives require evaluating the objective function at least $N+1$ times for $N$ parameters if one-sided finite differences are used, or $2N$ times for centered differences. As computing the objective function often involves solving a linear or nonlinear system, such as the MHD equilibrium equations, this implies solving the system of equations $\ge N+1$ times. Numerical derivatives also introduce additional noise, and the finite difference step size must be chosen carefully. To avoid these difficulties,
it is advantageous to compute shape gradients using adjoint methods \citep{Pironneau1974, Glowinski1975,Dekeyser2012,Dekeyser2014a,Dekeyser2014b}.
Adjoint methods allow the analytic derivative with respect to all $N$ parameters to be computed with only two solutions to the system of equations. Adjoint methods are thus much more efficient for computing derivatives with respect to many parameters, and they do not introduce the noise of numerical derivatives.
Adjoint methods were recently used in the context of stellarator design by 
 \cite{Paul2018}
for shape optimization of coil winding surfaces.

Rather than use parameter derivatives, in this work we will use an adjoint method to compute the shape gradient directly. This is sometimes termed adjoint shape sensitivity or adjoint shape optimization, which has its origins in aerodynamic engineering and computational fluid dynamics \citep{Pironneau1974,Glowinski1975}. As with adjoint methods for parameter derivatives, this technique only requires the solution of two linear or non-linear systems of equations. This technique has been applied to magnetic confinement fusion for the design of tokamak divertor shapes by solving forward and adjoint fluid edge equations \citep{Dekeyser2012,Dekeyser2014a,Dekeyser2014b}. As stellarators require many parameters to fully describe their shape, adjoint shape sensitivity could significantly decrease the cost of computing the shape gradient. If one is optimizing in the space of parameters describing the boundary of the plasma or the shape of coils, the shape gradient representation obtained from the adjoint method can be converted to parameter derivatives upon multiplication with a small matrix \citep{Landreman2018}. 

In \cref{sec:relations}, the fundamental adjoint relations for perturbations to MHD equilibria are derived and discussed. These relations take a form that is similar to that of transport coefficients that are related by Onsager symmetry \citep{Onsager1931}. Specifically, perturbations to the equilibrium are characterized as a set of generalized responses to a complementary set of generalized forces.  The responses and forces can be thought of as being related by a  matrix operator, which is symmetric.  The resulting relations among forces and responses can be used to compute the shape gradient of functions of the equilibria with respect to displacements of the plasma boundary, as in \cref{eq:shape_gradient_surface}, or the coil shapes, as in \cref{eq:shape_gradient_coil},  Several applications to stellarator figures of merit will be demonstrated in \cref{sec:applications}. While the primary application considered in this work will be stellarator optimization, the relations we obtain are equally applicable for 2D equilibria.

\section{Driven linear perturbations of 3D MHD equilibria}
\label{sec:relations}

The goal is to find a relation between small perturbations to a 3D magnetic confinement equilibrium configuration and the resulting change in a figure of merit of interest. As mentioned, the perturbations may be prescribed in one of two ways: either as a change in the shape of the outermost flux surface (fixed boundary) or as a change in the position, shape, or current strength in the coils confining the plasma (free boundary). 
(Even though the boundary shape changes in the former case, we refer to it as ``fixed boundary" since the equilibrium code is run in fixed boundary mode, and since the associated adjoint problem will turn out to have no boundary perturbation.) 
These perturbations will be referred to as the “true” perturbations and their direct calculation needs to be repeated many times for each possible change in shape of the outer flux surface or each change in the coil configuration to determine fully the sensitivity of the equilibrium. 

The approach we use is to instead calculate a different change in the equilibrium, which we refer to as the “adjoint” perturbation. The adjoint perturbation will correspond to the change in the equilibrium when an additional bulk force acts on the plasma, or the toroidal current profile is changed. For the adjoint perturbation there is no change to the outer flux surface in the fixed boundary case or to the coil currents in the free boundary case. In this section we will show that aspects of the true and adjoint changes are related to each other 
in a manner 
similar to Onsager symmetry. Thus, it will be shown that by calculating the adjoint perturbation, with a judiciously chosen added force or change in the toroidal current profile, the solution to the true problem can be determined. 

We consider equilibria in which the magnetic field in the plasma can be expressed in terms of scalar of functions $\psi(\textbf{x}),\Phi(\psi),\theta(\textbf{x})$, and $\zeta(\textbf{x})$,
\begin{gather}
\textbf{B} = \nabla \psi \times \nabla \theta - \nabla \Phi \times \nabla \zeta = \nabla \psi \times \nabla \alpha.
\label{eq:clebsch}
\end{gather}
We will regard $\psi$ as labeling the flux surfaces and consider toroidal geometries for which 
\begin{gather}
    \alpha = \theta - \iota(\psi) \zeta,
    \label{eq:field_line}
\end{gather}
label field lines in a flux surface, where $\theta$ is a poloidal angle, $\zeta$ is a toroidal angle, and $\iota(\psi) = d \Phi/d \psi$ is the rotational transform, with $\Phi$ being the poloidal flux function. (Any straight-field-line angles may be used.) 
With these definitions the magnetic flux passing toroidally through a poloidally closed curve of constant $\psi$ is $2\pi \psi$. The flux passing poloidally between the magnetic axis and the surface of constant $\psi$ is $2\pi \Phi(\psi)$. Thus, we assume that good flux surfaces exist and leave aside the issues of islands and chaotic field lines. 

In addition to the representation of the magnetic field, we assume that MHD force balance is satisfied with a scalar pressure, $p(\psi)$,
\begin{gather}
    0 =- \nabla p(\psi) + \frac{\textbf{J}\times\textbf{B}}{c},
    \label{eq:force_balance}
\end{gather}
where the current density, $\textbf{J}$, satisfies Ampere's law,
\begin{gather}
    \nabla \times \textbf{B} = \frac{4\pi}{c} \textbf{J},
    \label{eq:Ampere}
\end{gather}
and $c$ is the speed of light.
As mentioned, we will consider two cases, a fixed boundary case in which the shape of the outer flux surface is prescribed, and a free boundary case for which outside the plasma, whose surface is defined by a particular value of toroidal flux, the force balance equation (\ref{eq:force_balance}) does not apply, but rather, the magnetic field is determined by Ampere's law (\ref{eq:Ampere}) with a given current density $\textbf{J}_c$, representing current flowing in a set of coils.


From (\ref{eq:force_balance}) it follows that current density stream-lines also lie in the $\psi =$ constant surfaces. The toroidal current passing through a surface, $S_T(\psi)$, whose perimeter is a closed poloidal loop at constant $\psi$ is given by 
\begin{gather}
    I_T(\psi) = \int_{S_T(\psi)} d^2 x \, \textbf{n} \cdot \textbf{J} = \int_{S_T(\psi)} d \psi \, d \theta \, \sqrt{g} \nabla \zeta \cdot \textbf{J},
    \label{eq:toroidal_curr}
\end{gather}
where $\sqrt{g}^{-1} = \nabla \zeta \cdot \nabla \psi \times \nabla \theta$.

Equations \cref{eq:clebsch,eq:field_line,eq:field_line,eq:force_balance,eq:Ampere,eq:toroidal_curr} describe our base equilibrium configuration. We now consider small changes in the equilibrium that are assumed to yield a second equilibrium state of the same form as 
\cref{eq:clebsch},
but with new functions such that $\textbf{B}'=\nabla \psi' \times \nabla \theta' - \nabla \Phi'(\psi')\times \nabla \zeta'$. Each of the primed variables is assumed to differ from the corresponding unprimed variables by a small amount (e.g. $\psi' = \psi + \delta \psi(\textbf{x})$). The perturbed magnetic field can then be expressed $\textbf{B}'=\textbf{B}+\delta \textbf{B}$, where 
\begin{gather}
    \delta \textbf{B} = \nabla \delta \psi \times \nabla \theta + \nabla \psi \times \nabla \delta \theta - \nabla \Phi \times \nabla \delta \zeta - \nabla \left( \iota \delta \psi + \delta \Phi\right) \times \nabla \zeta.  
    \label{eq:delta_B}
\end{gather}
We write the perturbed poloidal flux as the sum of a term resulting from the perturbation of toroidal flux at fixed rotational transform, $\iota \delta \psi$, and a term representing the perturbed rotational transform, $\delta \Phi(\psi)$.  
Thus, we can regroup the terms in \cref{eq:delta_B} as follows
\begin{gather}
    \delta \textbf{B} = \nabla \times \left( \delta \psi \nabla \theta - \iota \delta \psi \nabla \zeta - \delta \theta \nabla \psi + \delta \zeta \nabla \Phi \right) - \nabla \delta \Phi(\psi) \times \nabla \zeta.
    \label{eq:delta_B2}
\end{gather}
The group of terms in parentheses 
in \cref{eq:delta_B2} corresponds to perturbations of the magnetic field allowed by ideal MHD, which is constrained by the ``frozen in law", and which preserves the rotational transform, ($\delta \iota =0$). The last term in \cref{eq:delta_B2} allows for changes in the rotational transform, ($\delta \iota = d \delta \Phi/d \psi$). 
Note also that the expression in parentheses in \cref{eq:delta_B2} can be written as a sum of terms parallel to $\nabla\psi$ and $\nabla\alpha$, and hence it is perpendicular to $\textbf{B}$. 
The group of terms in parentheses 
in \cref{eq:delta_B2} can thus be expressed in terms of a vector potential that is perpendicular to the equilibrium magnetic field, while the last term in \cref{eq:delta_B2} can be represented in terms of a vector potential in the toroidal direction, which thus has a component parallel to the equilibrium field. We can therefore write $\delta \textbf{B} = \nabla \times \delta \textbf{A}$, where
\begin{gather}
    \delta \textbf{A} = \bm{\xi} \times \textbf{B} - \delta \Phi \nabla \zeta. 
    \label{eq:delta_A1}
\end{gather}
Here, the variable $\bm{\xi}$ can be taken to be perpendicular to the applied magnetic field, and can be thought of as a displacement of the equilibrium magnetic field line. Using \cref{eq:delta_B} we write
\begin{gather}
    \bm{\xi} \times \textbf{B} = \delta \psi \left(\nabla \theta - \iota \nabla \zeta \right) - \delta \theta \nabla \psi + \delta \zeta \nabla \Phi,
    \label{eq:delta_A2}
\end{gather}
and from this we can see that perturbations of the toroidal and poloidal angles correspond to displacements in the flux surface, and the perturbation $\delta \psi$ gives a displacement out of the flux surface. A surface containing a prescribed toroidal flux $2\pi \psi_0$ in the unperturbed field is defined by $\psi(\textbf{x}) = \psi_0$. From \cref{eq:delta_A2}, 
the perturbation to $\psi$ at fixed position is $\delta \psi = - \bm{\xi} \cdot \nabla \psi$. When the field is perturbed the surface moves.
As $\bm{\xi}$ is a vector which describes the motion of field lines, the perturbed toroidal flux label moving with the $\psi$ surface, as measured in the unperturbed coordinate system, is given by $\psi(\textbf{x}) = \psi_0 + \bm{\xi} \cdot \nabla \psi(\textbf{x})$.

The change in toroidal current flowing through the perturbed surface is
\begin{gather}
    \delta I_T(\psi) = \int_{\partial S_T(\psi)} d \theta \, \sqrt{g}  \bm{\xi} \cdot \nabla \psi  \textbf{J} \cdot \nabla \zeta + \int_{S_T(\psi)} d \psi d \theta \, \sqrt{g}  \delta \textbf{J} \cdot \nabla \zeta, 
    \label{eq:delta_I}
\end{gather}
where $S_T(\psi)$ is a surface at constant toroidal angle bounded by the $\psi$ surface and $\partial S_T(\psi)$ is the boundary of such surface, a closed poloidal loop. Here the first term accounts for the displacement of the flux surface and the second term accounts for the change in toroidal current density. 

We now consider two distinct perturbations of the equilibrium of the type described by \cref{eq:delta_A1,eq:delta_A2,eq:delta_I}, which we denote with subscripts 1 and 2. In general, variables with subscript 1 will be associated with the true perturbation, and those with subscripts 2 will be associated with the adjoint perturbation. We then form the quantity
\begin{gather}
    U_T = \frac{1}{c}\int_{V_T} d^3 x \, \left( \delta \textbf{J}_1 \cdot \delta \textbf{A}_2 - \delta \textbf{J}_2 \cdot \delta \textbf{A}_1 \right) = 0, 
    \label{eq:Ut}
\end{gather}
where the variables $\delta \textbf{J}$ and $\delta \textbf{A}$ are the changes in current density and vector potential associated with the two perturbations, and the integral is, for the time being, over all space. That $U_T = 0$ follows from expressing the change in current densities in terms of the change in magnetic fields via Ampere's law. This turns the integrand in \cref{eq:Ut} into a divergence, which in turn becomes a surface integral, which we take to be at infinity where fields vanish sufficiently fast. 

We now express the volume integral in \cref{eq:Ut} as the sum of three terms,
\begin{gather}
    U_T = U_P + U_B + U_C = 0.
    \label{UT}
\end{gather}
Here $U_P$ is the contribution from volume in the plasma, integrated just up to the plasma-vacuum boundary. For this term we represent the vector potentials using \cref{eq:delta_A1}
\begin{gather}
    U_P = \frac{1}{c} \int_{V_P} d^3 x \, \left( \delta \textbf{J}_1 \cdot \left( \bm{\xi}_2 \times \textbf{B} - \delta \Phi_2 \nabla \zeta \right) - \delta \textbf{J}_2 \cdot \left( \bm{\xi}_1 \times \textbf{B} - \delta \Phi_1 \nabla \zeta \right) \right).
    \label{eq:Upa}
\end{gather}

To evaluate \cref{eq:Upa} we use the perturbed force balance relation
\begin{gather}
    0 =  \delta\textbf{F} +\nabla(\bm{\xi}\cdot\nabla p)+ \frac{\delta \textbf{J} \times \textbf{B} + \textbf{J} \times \delta \textbf{B}}{c},
    \label{eq:perturbed_force_balance}
\end{gather}
where $\delta\textbf{F}$ is an additional perturbed force to be prescribed. The term $\nabla(\bm{\xi}\cdot\nabla p)$ represents the perturbed force associated with the perturbed pressure in the case of a true solution for which the dependence of the pressure on flux is preserved. As the perturbation to the flux label at fixed position is $\delta \psi = - \bm{\xi} \cdot \nabla \psi$ and the pressure is assumed to be a fixed function $p(\psi)$, then the perturbation to the pressure at fixed position is $\delta p = - \bm{\xi} \cdot \nabla p(\psi)$. If the equilibrium calculation is performed with specified $p(\psi)$, this does not imply any additional restrictions.

The term $U_B$ comes from integrating over a thin layer at the plasma-vacuum boundary. At the boundary the difference between the perturbed and unperturbed current density has the character of a current sheet due to the displacement of the outermost flux surface. This effective current sheet causes a jump in the tangential components of the perturbation to the magnetic fields at the surface. This jump implies that care must be taken in evaluating the perturbed magnetic fields at the surface as they have different values on either side of the plasma-vacuum surface. However, the vector potential is continuous at the plasma-vacuum boundary. Thus, we write
\begin{gather}
    U_B = \frac{1}{c} \int_{S_p} \frac{d^2 x}{| \nabla \psi |} \left( \bm{\xi}_1 \cdot \nabla \psi \textbf{J} \cdot \delta \textbf{A}_2 - \bm{\xi}_2 \cdot \nabla \psi \textbf{J} \cdot \delta \textbf{A}_1 \right), 
\end{gather}
where the vector potentials are expressed as in \cref{eq:delta_A1}. Using this expression for the vector potentials and expressing the surface integral as an integral over the toroidal and poloidal angles gives
\begin{gather}
    U_B = \frac{1}{c} \int_{S_p} d \theta d \zeta \, \sqrt{g} \textbf{J} \cdot \nabla \zeta \left( - \bm{\xi}_1 \cdot \nabla \psi \delta \Phi_2 + \bm{\xi}_2 \cdot \nabla \psi \delta \Phi_1 \right). 
    \label{eq:Ub}
\end{gather}
Here we note the terms in the vector potential coming from the MHD displacement cancel. 

Last, the quantity $U_C$ represents the contribution from the integral over the volume outside the plasma where only the coil currents need to be included 
\begin{gather}
    U_C = \frac{1}{c} \int_{V_V} d^3 x \, \left( \delta \textbf{J}_{C_1} \cdot \delta \textbf{A}_{V_2} - \delta \textbf{J}_{C_2} \cdot \delta \textbf{A}_{V_1} \right),
    \label{eq:Uc}
\end{gather}
where $\delta \textbf{A}_{V_{1,2}}$ is the change in the vacuum vector potential, and $\delta \textbf{J}_{C_{1,2}}$ is the change in the coil current density. 

Combining $U_P$, $U_B$, and $U_C$ gives the following relation appropriate to the free boundary case $U_T = U_P + U_B + U_C = 0$, or
\begin{multline}
    \int_{V_P} d^3 x \, \left(- \bm{\xi}_1 \cdot  \delta \textbf{F}_2   + \bm{\xi}_2 \cdot  \delta \textbf{F}_1 \right) + \frac{2\pi}{c} \int_{V_P} d \psi \left( \delta \Phi_1 \der{\delta I_{T,2}}{\psi}-\delta \Phi_2 \der{\delta I_{T,1}}{\psi} \right) \\ + \frac{1}{c} \int_{V_V} d^3 x \, \left( \delta \textbf{J}_{C_1} \cdot \delta \textbf{A}_{V_2} - \delta \textbf{J}_{C_2} \cdot \delta \textbf{A}_{V_1} \right) = 0.
    \label{eq:free_boundary}
\end{multline}
The many steps leading to \cref{eq:free_boundary} are outlined in \cref{app:adjoint_relation}.

A similar relation can be obtained in the fixed boundary case. Here the integral over the plasma volume, \eqref{eq:Upa}, can be written as a surface integral by applying the divergence theorem,
\begin{gather}
    U_P = \frac{1}{4\pi}\int_{S_P} d^2 x \, \textbf{n} \cdot \left( \delta \textbf{B}_1 \times \delta \textbf{A}_2 - \delta \textbf{B}_2 \times \delta \textbf{A}_1 \right).
    \label{eq:Up_bound}
\end{gather}
Again, following steps outlined in \cref{app:adjoint_relation}, this may be rewritten in the following form, 
\begin{multline}
    \int_{V_P} d^3 x \, \left(- \bm{\xi}_1 \cdot  \delta \textbf{F}_2   + \bm{\xi}_2 \cdot  \delta \textbf{F}_1 \right)
    - \frac{2\pi}{c} \int_{V_P} d \psi \, \left( \delta I_{T,2} \der{\delta \Phi_1}{\psi} - \delta I_{T,1} \der{\delta \Phi_2}{\psi} \right) \\
    - \frac{1}{4\pi} \int_{S_P} d^2 x \, \textbf{n} \cdot \left( \bm{\xi}_2 \delta \textbf{B}_1 \cdot \textbf{B} - \bm{\xi}_1 \delta \textbf{B}_2 \cdot \textbf{B} \right) = 0. 
    \label{eq:fixed_boundary}
\end{multline}
The fixed boundary adjoint relation can also be obtained by applying the self-adjointness of the MHD force operator (\cref{appendix:self-adjointness}). If the second term in \eqref{eq:fixed_boundary} is integrated by parts in $\psi$, we see that the fixed and free boundary adjoint relations share the terms involving the products of displacements with bulk forces and perturbed fluxes with perturbed toroidal currents. The integral over the vacuum region in \eqref{eq:free_boundary} is replaced by an integral over the plasma boundary and a boundary term from the integration by parts in $\psi$ in \eqref{eq:fixed_boundary}. 


We now have two integral relations between perturbations 1 and 2, viz \cref{eq:free_boundary,eq:fixed_boundary}. They have a common form in that they each are the sum of three integrals: the first involving 
forces 
and displacements, the second involving the toroidal current and poloidal flux profiles, and the third involving the manner in which the plasma boundary is prescribed. In \cref{eq:free_boundary}, the free boundary case, the changes in coil current densities are specified. In \cref{eq:fixed_boundary}, the fixed boundary case, the displacement of the outer flux surface is prescribed. Equations \cref{eq:free_boundary,eq:fixed_boundary} can also be viewed as the difference in sums of generalized forces and responses.  For example, in \cref{eq:free_boundary} we can consider the quantities $\delta \textbf{F}$, $\delta \Phi$, $\delta \textbf{J}_c$ as forces and $\bm{\xi}$, $d \delta I_T/d\psi$, $\delta \textbf{A}_V$ as responses.  The fact that the sum of the products of true forces and adjoint responses less the products of adjoint forces and true responses vanishes is similar to the relation between forces and fluxes related by Onsager symmetry.  In the case of Onsager symmetry this relation follows from the self-adjoint property of the collision operator.  In the case of MHD equilibria it is known that the force operator is self-adjoint.  Here we see that the self-adjoint property is extended to cases in which the MHD frozen-in constraint is broken.

These relations \cref{eq:free_boundary} and \cref{eq:fixed_boundary} can be used to generate the shape gradient if we manipulate the terms to be zero, or some known quantity. For example, if we impose that both the true and adjoint solutions maintain pressure as a function of flux, or equivalently there are no added forces, then the first terms in both \cref{eq:free_boundary} and \cref{eq:fixed_boundary} vanish. If we impose that 
the adjoint solution in \cref{eq:fixed_boundary} involves no change in shape of the outer flux surface, the boundary term involving $\textbf{n} \cdot \bm{\xi}_2$ vanishes. 
This leaves 
\begin{gather}
    \frac{2\pi}{c} \int_{V_P} d \psi \left( \delta I_{T,2} \der{\delta \Phi_1}{\psi} - \delta I_{T,1} \der{\delta \Phi_2}{\psi} \right) = \frac{1}{4\pi} \int_{S_p}d^2 x \, \textbf{n} \cdot \bm{\xi}_1 \delta \textbf{B}_2 \cdot \textbf{B} .
\end{gather}
Thus, if we solve the adjoint problem with a prescribed change in toroidal current profile, and ask what is the change in rotational transform in the true problem when the current profile is unchanged we find
\begin{gather}
    \frac{2\pi}{c} \int_{V_P} d \psi \left( \delta I_{T,2} \der{\delta \Phi_1}{\psi} \right) = \frac{1}{4\pi}\int_{S_P} d^2 x \, \mathbf{n} \cdot \bm{\xi}_1 \delta \textbf{B}_2 \cdot \textbf{B} . 
\end{gather}
The right-hand side has the form of \cref{eq:shape_gradient_surface}, with $\delta \textbf{B}_2\cdot\textbf{B}/4\pi$ playing the role of the shape gradient. 
Thus, by picking an adjoint current profile change that is localized to a particular flux surface, we determine the sensitivity of the rotational transform at that flux surface to changes in the boundary shape. Note that even if the rotational transform is allowed to vary, the normal component of $\bm{\xi}_1$ describes the displacement of the boundary, as discussed in \cref{app:displacement}. Thus the form of the shape gradient in \eqref{eq:shape_gradient_surface} can be used with $\delta \textbf{r} \cdot \textbf{n}$ replaced with $\bm{\xi}_1 \cdot \textbf{n}$. 

If we impose that the adjoint solution in \eqref{eq:free_boundary} involves no change in coil currents, upon integrating the middle term in \cref{eq:free_boundary} by parts a similar relation can be obtained for the free boundary case, 
\begin{gather}
\frac{2\pi}{c} \int_{V_P} d \psi \, \left( \delta I_{T,2} \der{\delta \Phi_1}{\psi} \right) = \frac{1}{c} \int_{V_V} d^3 x \, \delta \textbf{J}_{C_1} \cdot \delta \textbf{A}_{V_2},
\end{gather}
where it has been assumed that $\delta I_{T,2} = 0$ at the boundary. 

When the coil currents are confined to wires, the right hand side can be expressed as changes in currents and fluxes and integrals along wires, 
\begin{gather}
    \frac{1}{c} \int_{V_V}d^3 x \,  \delta \textbf{J}_{C_1} \cdot \delta \textbf{A}_{V_2}  = \frac{1}{c} \sum_{k} \left(\delta \Phi_{C_{2,k}} \delta I_{C_{1,k}} + I_{C_k} \int_{C_k} dl \, \delta \textbf{r}_{C_k}(\textbf{x}) \cdot \textbf{t} \times \delta \textbf{B}_2 \right).
    \label{eq:coil_perturb}
\end{gather}
Here $\delta \Phi_{C_{2,k}}$, $\delta I_{C_{1,k}}$ are the change in adjoint flux through and change in true current in coil $k$, and $I_{C_k}$ is the current through the unperturbed coil. The unit tangent vector along $C_k$ is $\textbf{t}$. The second term describes the effect of moving the coil. It is proportional to the current in the coil and the line integral of the displacement of the coil dotted with the cross product of the tangent to the coil and the perturbed adjoint magnetic field. 

The sensitivity of other figures of merit may be evaluated by considering cases for which both the true and adjoint solutions preserve the rotational transform or the toroidal current profile. In this case a perturbed force density 
is included in the adjoint calculation
\begin{gather}
   \int_{V_P} d^3 x \, \bm{\xi}_1 \cdot  \delta \textbf{F}_2  = \frac{1}{c} \int_{V_V} d^3 x \,  \delta \textbf{J}_{C_1} \cdot \delta \textbf{A}_{V_2}. 
\end{gather}
In this case one must find an expression for the perturbed adjoint force, $\delta \textbf{F}_2$, such that the left hand side is the change in a figure of merit of interest. Some examples will be discussed in the next section. 

\section{Applications}
\label{sec:applications}

In this section we will consider figures of merit which depend on the shape of the outer boundary of the plasma (\cref{sec:surf_Beta} and  \ref{sec:surf_iota}) and on the shape of the electromagnetic coils (\cref{sec:coil_iota}). The shape gradients of these figures of merit will be computed using both a direct method and an adjoint method,
to demonstrate that the adjoint method produces identical results to the direct method but at much lower computational expense. 

\subsection{Surface shape gradient for $\beta$}
\label{sec:surf_Beta}

Consider a figure of merit, the volume-averaged $\beta$,
\begin{align}
f_{\beta} = \frac{f_P}{f_B}
\label{eq:beta},
\end{align}
where 
\begin{align}
    f_{P} = \int_{V_p} d^3 x \, p(\psi)
\end{align}
and 
\begin{align}
    f_B = \int_{V_p} d^3 x \, \frac{B^2}{8\pi}.
\end{align}
(This definition of volume-averaged $\beta$ is the one employed in the VMEC code \citep{Hirshman1983}.)  
While $f_{\beta}$ is a figure of merit not often considered in stellarator shape optimization, we include this calculation to demonstrate the adjoint approach, as its shape gradient can be computed without modifications to an equilibrium code.

The differential change in $f_P$ associated with displacement $\bm{\xi}_1$ is 
\begin{gather}
    \delta f_P(S_P;\bm{\xi}_1) = -\int_{V_P} d^3 x \, \bm{\xi}_1 \cdot \nabla p + \int_{S_P} d^2 x \, \bm{\xi}_1 \cdot \textbf{n} p(\psi).
\end{gather}
The first term accounts for the change in $p$ at fixed position due to the motion of the flux surfaces, and the second term accounts for the motion of the boundary. The differential change in $f_B$ associated with $\bm{\xi}_1$ is
\begin{multline}
    \delta f_B(S_P;\bm{\xi}_1) = -\frac{1}{4\pi}\int_{V_P} d^3 x \, \left( B^2 \nabla \cdot \bm{\xi}_1 + \bm{\xi}_1 \cdot \nabla \left( B^2 + 4\pi p \right) \right) \\ + \frac{1}{8\pi}\int_{S_P} d^2 x \, \bm{\xi}_1 \cdot \textbf{n} B^2.
    \label{eq:df_B}
\end{multline}
where we have assumed a perturbation that preserves the rotational transform ($\delta \Phi_1 = 0$), for which $\delta B^2 = - 2\left(B^2 \nabla \cdot \bm{\xi}_1 + \bm{\xi}_1 \cdot \nabla \left( B^2 + 4\pi p\right) \right)$ is the perturbation to the magnetic field strength at fixed position. The first term in \eqref{eq:df_B} corresponds with the change in $f_B$ due to the perturbation to the field strength, while the second term accounts for the motion of the boundary. Applying the divergence theorem we obtain,
\begin{gather}
    \delta f_B(S_P;\bm{\xi}_1) = -\int_{V_P} d^3 x \, \bm{\xi}_1 \cdot \nabla p - \frac{1}{8\pi}\int_{S_P} d^2 x \,   \bm{\xi}_1 \cdot \textbf{n} B^2. 
\end{gather}

The differential change in $f_{\beta}$ associated with displacement $\bm{\xi}_1$ satisfies
\begin{multline}
    \frac{\delta f_{\beta}(S_p;\bm{\xi}_1)}{f_{\beta}} =  \int_{S_P} d^2 x \, \bm{\xi}_1 \cdot \textbf{n} \left(\frac{p(\psi)}{f_P} +  \frac{B^2}{8\pi f_B}\right)
    -\left( \frac{1}{f_P} - \frac{1}{f_B} \right) \int_{V_P} d^3 x \, \bm{\xi}_1 \cdot \nabla p.
    \label{eq:f_Beta1}
\end{multline}
The first term on the right of \cref{eq:f_Beta1} is already in the form of a shape gradient. To evaluate the second term, we turn to the adjoint problem, and we choose $\delta \textbf{F}_2 = -\nabla \left( \Delta p\right)$, where $\Delta \ll 1$ is a constant scalar. That is, we add a force which is proportional to the equilibrium pressure force with a small multiplier, $\Delta$. This additional force produces a proportional change in magnetic field at the boundary and thus from \cref{eq:fixed_boundary}, we find
\begin{gather}
    \frac{\delta f_{\beta}(S_P;\bm{\xi}_1)}{f_{\beta}} = \int_{S_P} d^2 x \, \bm{\xi}_1 \cdot \textbf{n} \left( \frac{p(\psi)}{f_P} + \frac{B^2}{8\pi f_B} + \left( \frac{1}{f_P} - \frac{1}{f_B}  \right) \frac{\delta \textbf{B}_2 \cdot \textbf{B} }{4\pi \Delta } \right). 
\end{gather}
Thus, we can obtain the shape gradient without perturbing the shape of the surface,
\begin{gather}
    S = f_{\beta} \left( \frac{p(\psi)}{f_P} + \frac{B^2}{8\pi f_B} + \left( \frac{1}{f_P} - \frac{1}{f_B}  \right) \frac{\delta \textbf{B}_2 \cdot \textbf{B} }{4\pi \Delta } \right). 
    \label{eq:beta_adjoint}
\end{gather}
Obtaining $S$ amounts to computing an equilibrium with unperturbed $\iota$, unperturbed boundary, and perturbed pressure $p' = (1 + \Delta)p$. 

A similar expression can be obtained for equilibria for which the rotational transform is allowed to vary, but the toroidal current is held fixed ($\delta I_{T,1} = 0$). In this case, the toroidal current for the adjoint problem is chosen to be $\delta I_{T,2} = -\Delta I_T \left( 1/f_P - 1/f_B\right)^{-1} \left(1/f_B \right)$ where $I_T$ is the unperturbed current profile, and again $\delta \textbf{F}_2 = -\nabla \left( \Delta p \right)$. The shape gradient can then be obtained from \eqref{eq:beta_adjoint}. 

\begin{figure}
    \centering
    \begin{subfigure}[b]{0.8\textwidth}
    \centering
    \includegraphics[trim ={1.5cm 8cm 1.5cm 10cm},clip,width=1.0\textwidth]{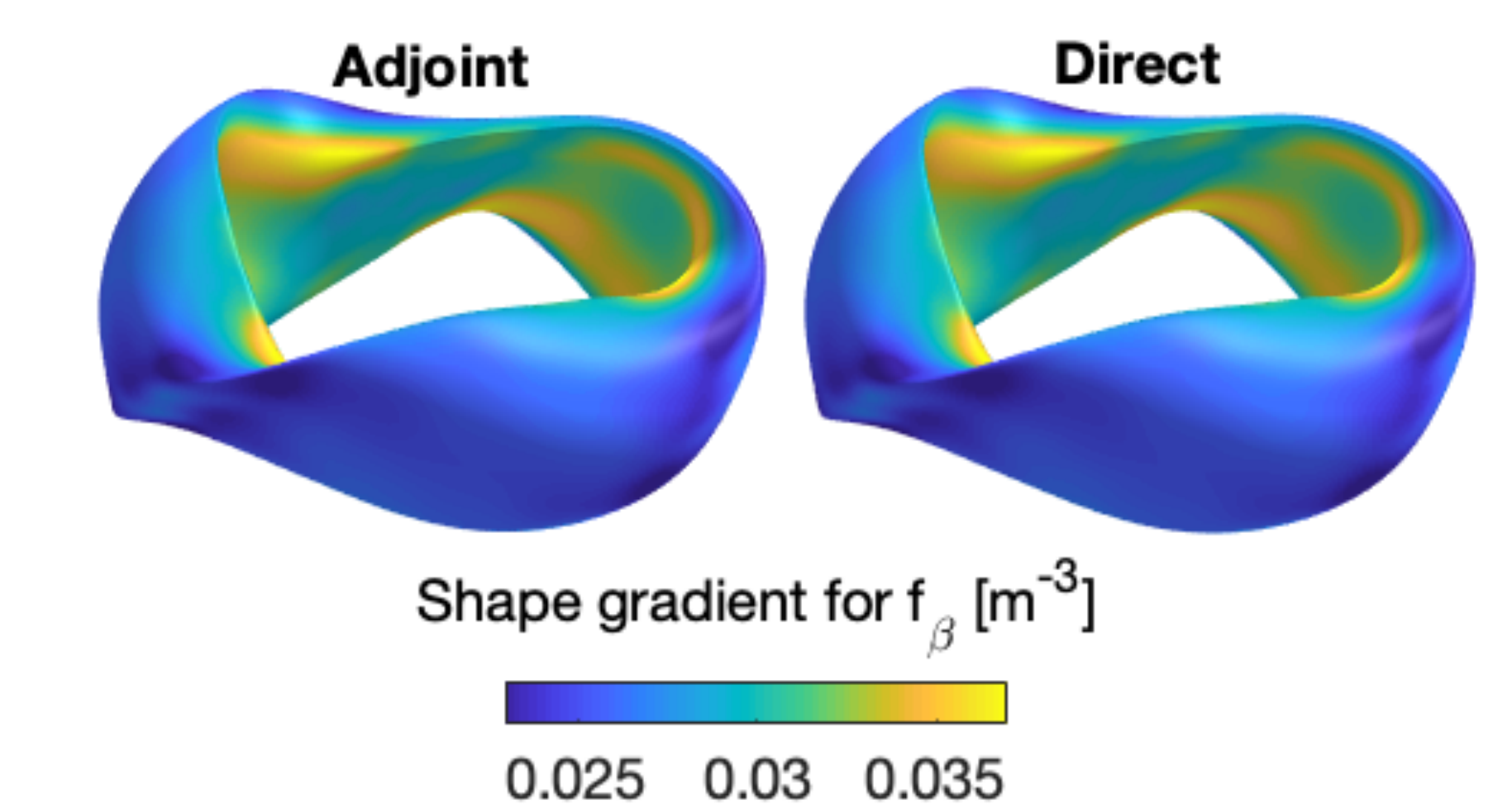}
    \caption{}
    \label{fig:beta_torus}
    \end{subfigure}
    \begin{subfigure}[b]{0.49\textwidth}
    \centering
    \includegraphics[trim={1cm 6cm 1cm 6.5cm},clip,width=1.0\textwidth]{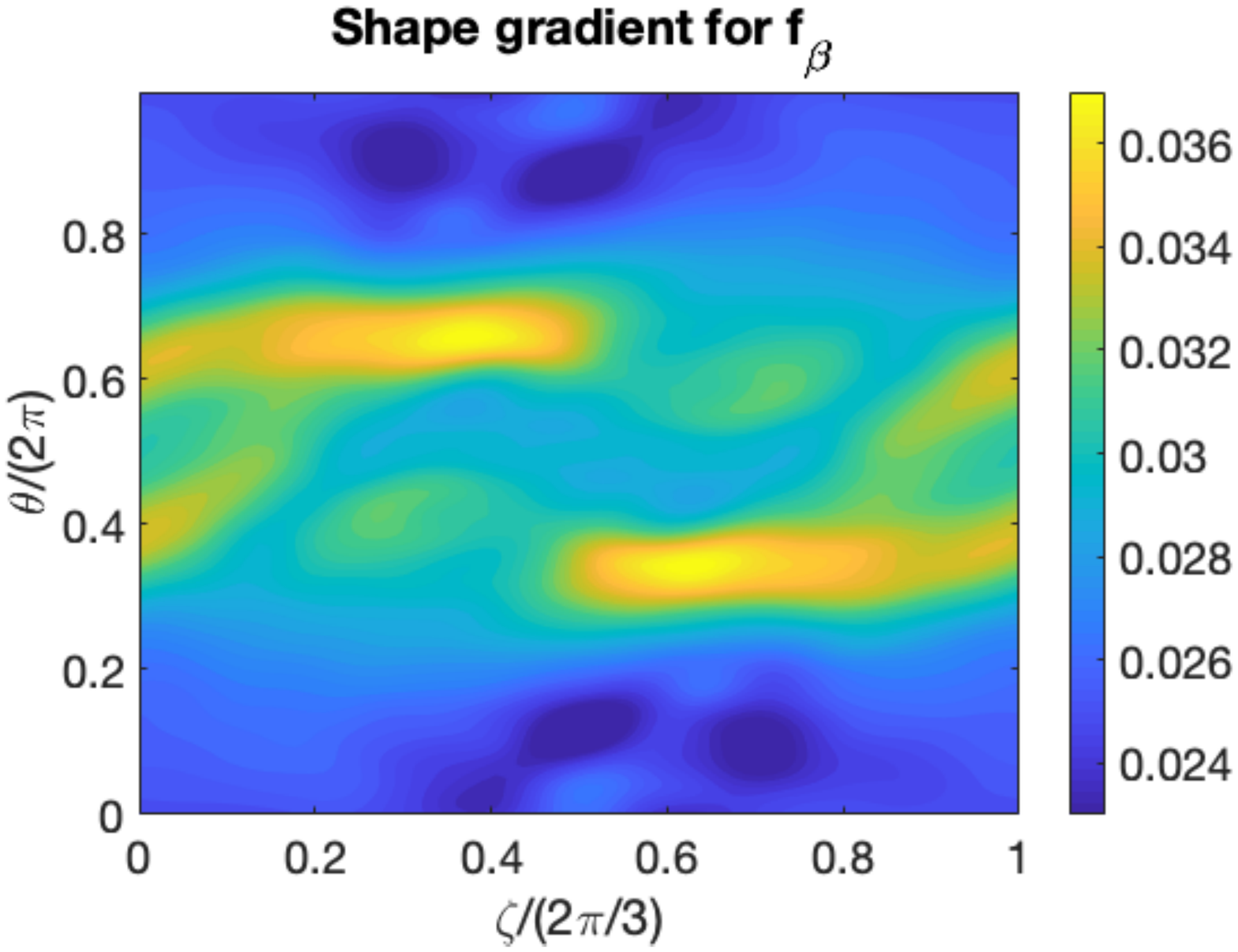}
    \caption{}
    \end{subfigure}
    \begin{subfigure}[b]{0.48\textwidth}
    \centering
    \includegraphics[trim={1cm 6cm 1cm 6.5cm},clip,width=1.0\textwidth]{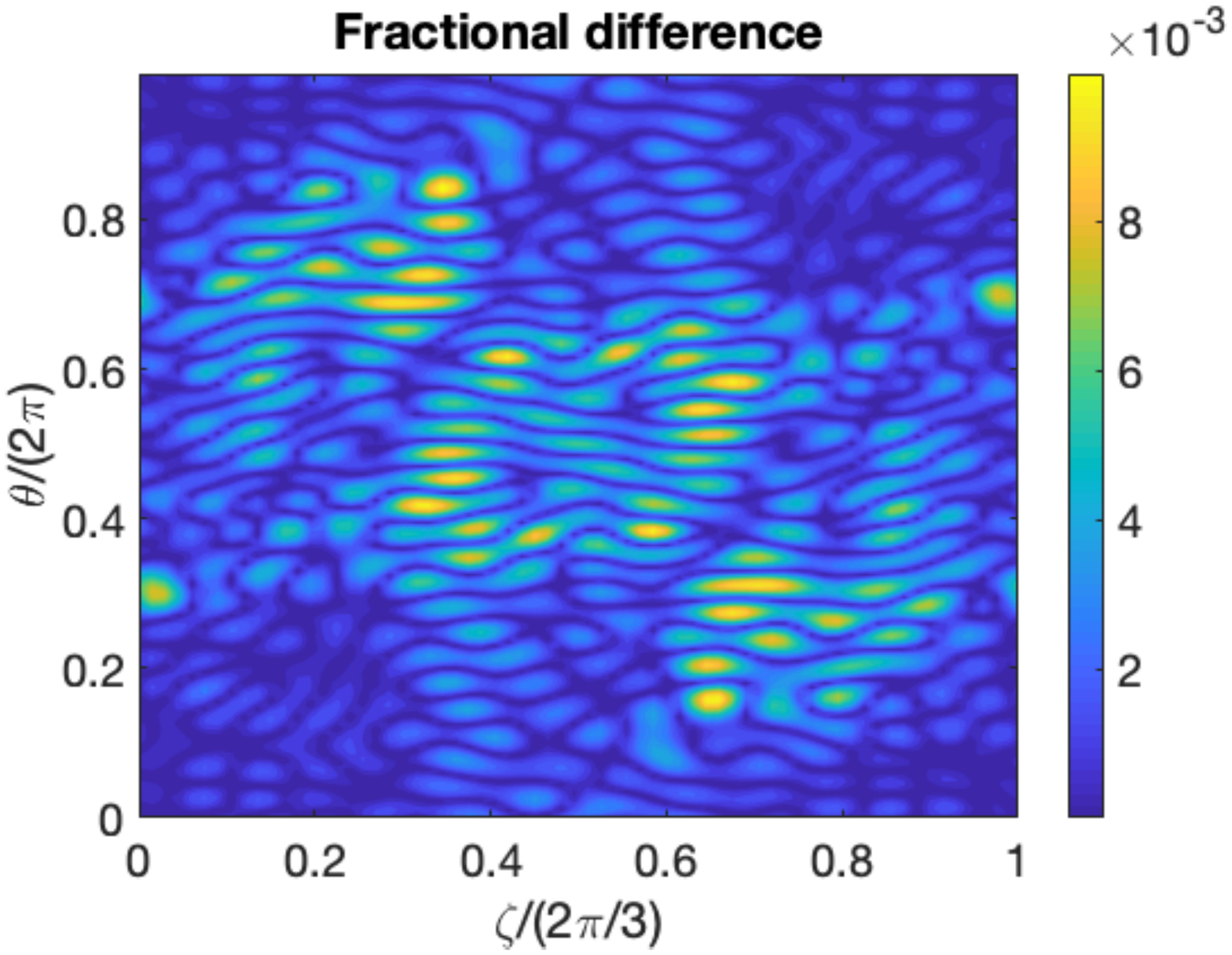}
    \caption{}
    \label{fig:beta_residual}
    \end{subfigure}
    \begin{subfigure}[b]{0.48\textwidth}
    \centering
    \includegraphics[trim={1cm 6cm 1cm 7cm},clip,width=1.0\textwidth]{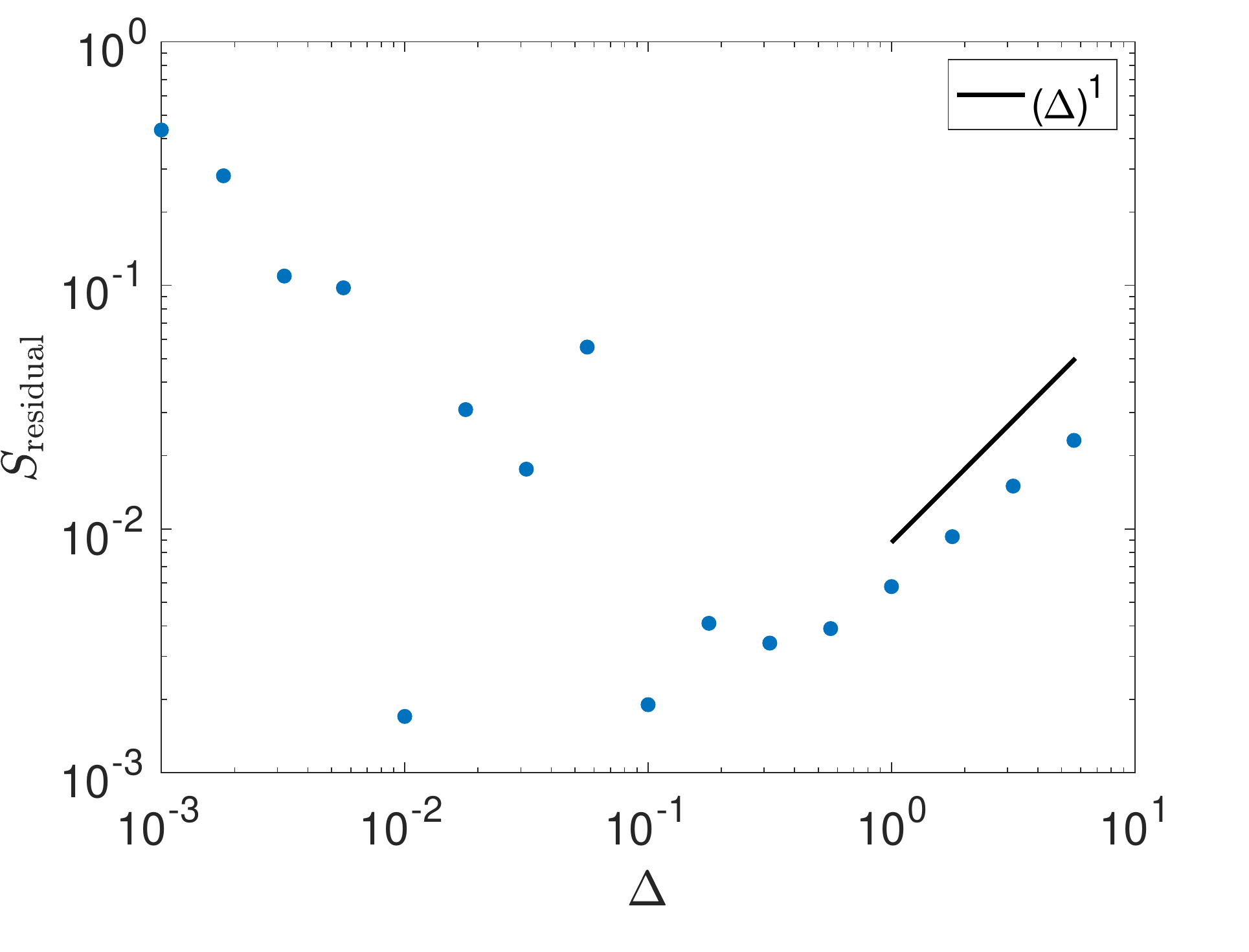}
    \caption{}
    \label{fig:delta_scan}
    \end{subfigure}
    \caption{(a) The shape gradient for $f_{\beta}$ \cref{eq:beta} computed using the adjoint solution \cref{eq:beta_adjoint} (left) and using parameter derivatives (right). (b) The shape gradient computed with the adjoint solution in the $\zeta-\theta$ plane. (c) The fractional difference \eqref{eq:residual} between the shape gradient obtained with the adjoint solution and with parameter derivatives. The two methods give virtually indistinguishable results, as they should. (d) The fractional difference between the shape gradient obtained with the adjoint solution and with parameter derivatives, $S_{\text{residual}}$, depends on the scale of the perturbation added to the adjoint force balance equation, $\Delta$.}
    \label{fig:beta_shape_gradient}
\end{figure}

To demonstrate, we use the NCSX stellarator LI383 equilibrium \citep{Zarnstorff2001}. The pressure profile was perturbed with $\Delta = 0.01$. The unperturbed and adjoint equilibria are computed with the VMEC code \citep{Hirshman1983}. The shape gradient obtained with the adjoint solution, $S_{\text{adjoint}}$, and that obtained with the direct approach, $S_{\text{direct}}$, are shown in \cref{fig:beta_torus}. Positive values of the shape gradient indicate that $f_{\beta}$ increases if a normal perturbation is applied at a given location as indicated by \eqref{eq:shape_gradient_surface}. For the direct approach, parameter derivatives $(\partial f_{\beta}/\partial R_{mn}^c, \partial f_{\beta}/\partial Z_{mn}^s)$ are computed with a centered 4-point stencil for $m \leq 15$ and $|n| \leq 9$ using a polynomial fitting technique. The shape gradient is obtained using the method outlined in section 4.2 of \cite{Landreman2018}. The fractional difference between the two methods,
\begin{gather}
S_{\text{residual}} = \frac{|S_{\text{adjoint}}-S_{\text{direct}}|}{\sqrt{\int_{S_P} d^2 x \, S_{\text{adjoint}}^2/\int_{S_P} d^2 x }},
\label{eq:residual}
\end{gather}
is shown in \cref{fig:beta_residual}. The surface-averaged value of $S_{\text{residual}}$ is $1.7\times10^{-3}$.

The parameter $\Delta$ must be chosen carefully, as the perturbation must be large enough that the result is not dominated by round-off error, but small enough that non-linear effects do not become important. The relationship between $S_{\text{residual}}$ and $\Delta$ is shown in figure \ref{fig:delta_scan}. Here $S_{\text{direct}}$ is computed using the parameters reported above such that convergence is obtained. We find that $S_{\text{residual}}$ decreases as $\left(\Delta\right)^1$ until $\Delta \approx 0.5$, at which point round-off error begins to dominate. This scaling is to be expected, as $\delta \textbf{B}_2$ is computed with a forward difference derivative with step size $\Delta$. 

For this and the later examples, the computational cost of transforming the parameter derivatives to the shape gradient was negligible compared to the cost of computing the parameter derivatives. 
The direct approach used 2357 calls to VMEC while the adjoint approach only required two. 
It is clear that the adjoint method yields nearly identical derivative information to the direct method but at substantially reduced computational cost. 

In \cref{fig:beta_shape_gradient} we find that the shape gradient for $f_{\beta}$ is everywhere positive. This reflects the fact that the toroidal flux enclosed by $S_P$ is fixed. As perturbations which displace the plasma surface outward increase the surface area of a toroidal cross-section, the toroidal field must correspondingly decrease, thus increasing $f_{\beta}$. We find that the shape gradient is increased in regions of large field strength, as indicated by the second term in \eqref{eq:beta_adjoint}.

\subsection{Surface shape gradient for rotational transform}
\label{sec:surf_iota}


Consider a figure of merit, the average rotational transform in a radially localized region,
\begin{gather}
    f_{\iota} = \int_{V_P} d \psi \, \iota(\psi) w(\psi).
    \label{eq:iota}
\end{gather}
Here $w(\psi)$ is a normalized weighting function,
\begin{gather}
    w(\psi) = \frac{e^{-(\psi-\psi_m)^2/\psi_w^2}}{\int_{V_P} d \psi \, e^{-(\psi-\psi_m)^2/\psi_w^2}},
    \label{eq:weighting}
\end{gather}
and $\psi_m$ and $\psi_w$ are parameters defining the center and width of the Gaussian weighting, respectively.

The differential change of $f_{\iota}$ associated with perturbation $\bm{\xi}_1$ is
\begin{gather}
    \delta f_{\iota}(S_p;\bm{\xi}_1) =  \int_{V_P} d \psi \, \der{\delta \Phi_1}{\psi} w(\psi).
    \label{eq:iota_deriv}
\end{gather}
For the adjoint problem, we can choose a toroidal current profile $\delta I_{T,2} = I_{\Delta} w(\psi)$, where $I_{\Delta}$ is a scalar constant, and we can take $\delta \textbf{F}_2 = 0$. This additional current produces a proportional change in the magnetic field at the boundary; thus using \cref{eq:fixed_boundary}, we obtain the following
\begin{gather}
    \delta f_{\iota}(S_p;\bm{\xi}_1) = \frac{c}{I_{\Delta} 8 \pi^2} \int_{S_P}d^2 x \, \textbf{n} \cdot \bm{\xi}_1 \delta \textbf{B}_2 \cdot \textbf{B}.
\end{gather}
So, we can obtain the shape gradient from the adjoint solution
\begin{gather}
    S = \frac{c\delta \textbf{B}_2 \cdot \textbf{B}}{I_{\Delta} 8\pi^2}.
    \label{eq:iota_adjoint}
\end{gather}
Note that the computation of the shape derivative of the rotational transform on a single surface, $\psi_m$, with the adjoint approach would require a delta-function current perturbation, $\delta I_{T,2} = I_{\Delta} \delta (\psi-\psi_m)$. As this type of perturbation is difficult to resolve in a numerical computation, the use of the Gaussian envelope allows the shape derivative of the rotational transform in a localized region of $\psi_m$ to be computed. 

\begin{figure}
    \centering
    \begin{subfigure}[b]{0.8\textwidth}
    \centering
    \includegraphics[trim ={1.5cm 8cm 1.5cm 10cm},clip,width=1.0\textwidth]{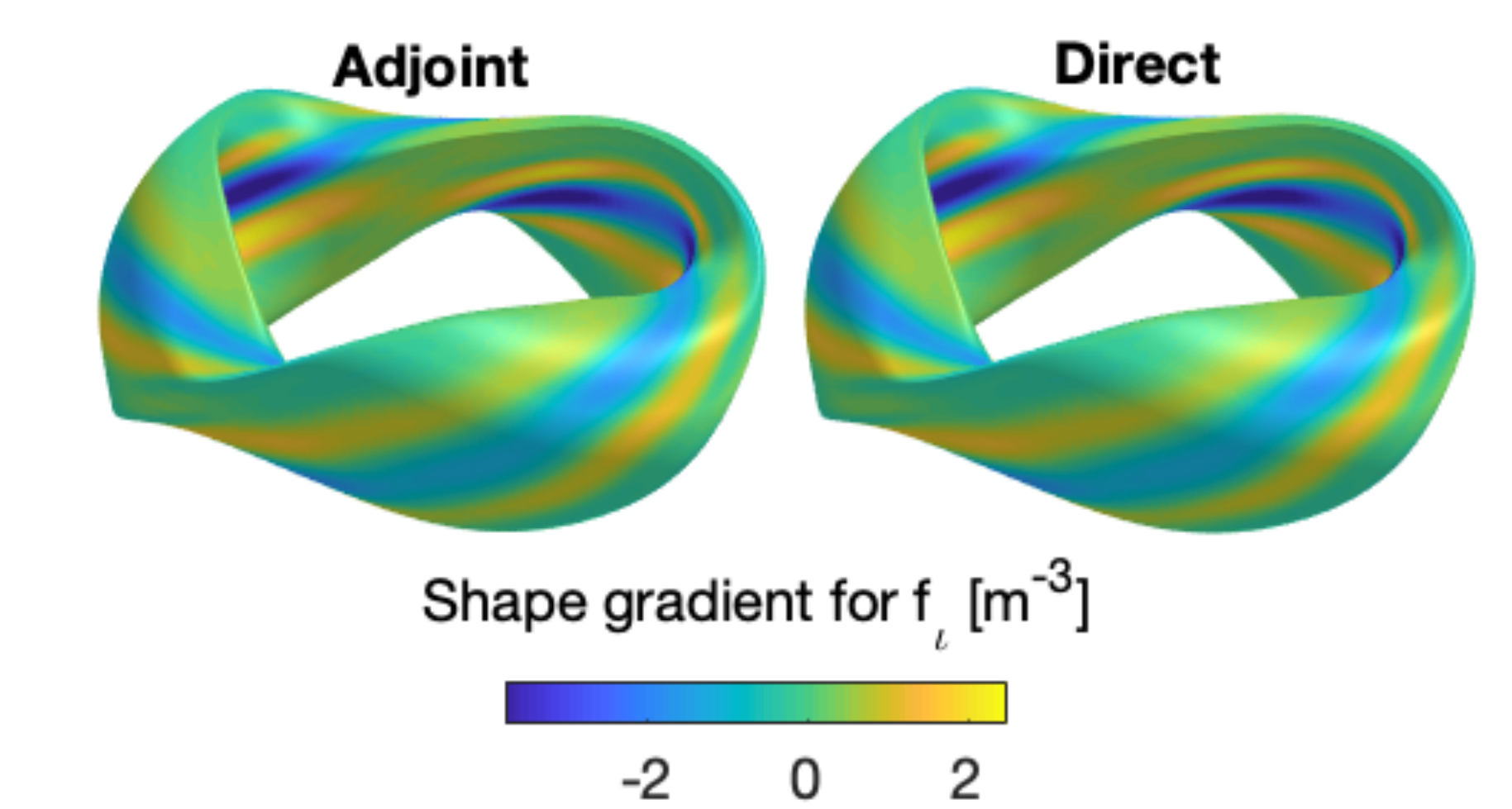}
    \caption{}
    \label{fig:iota_torus}
    \end{subfigure}
    \begin{subfigure}[b]{0.49\textwidth}
    \includegraphics[trim={1cm 6cm 1cm 6cm},clip,width=1.0\textwidth]{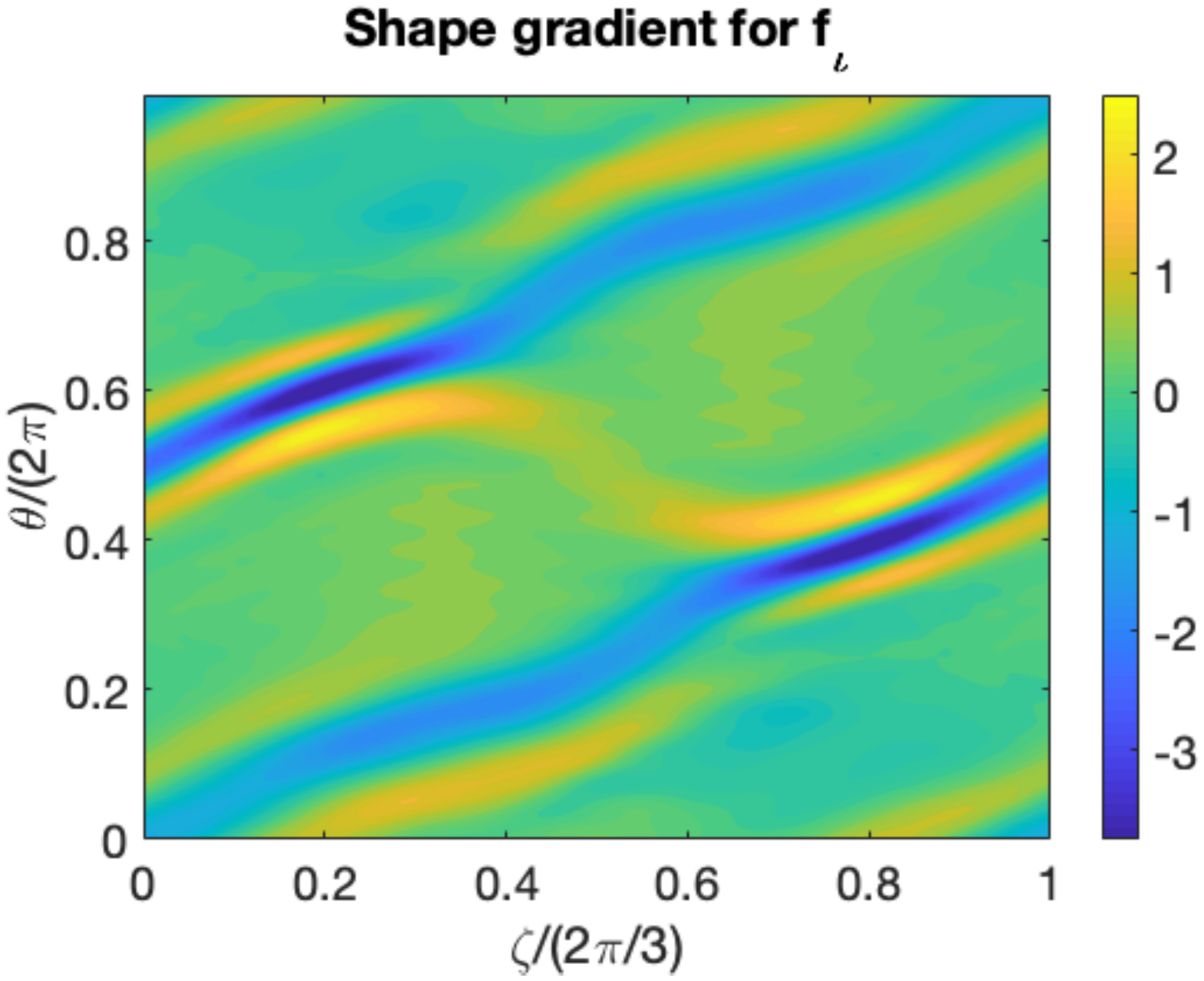}
    \caption{}
    \end{subfigure}
    \begin{subfigure}[b]{0.48\textwidth}
    \includegraphics[trim={1cm 6cm 1cm 6cm},clip,width=1.0\textwidth]{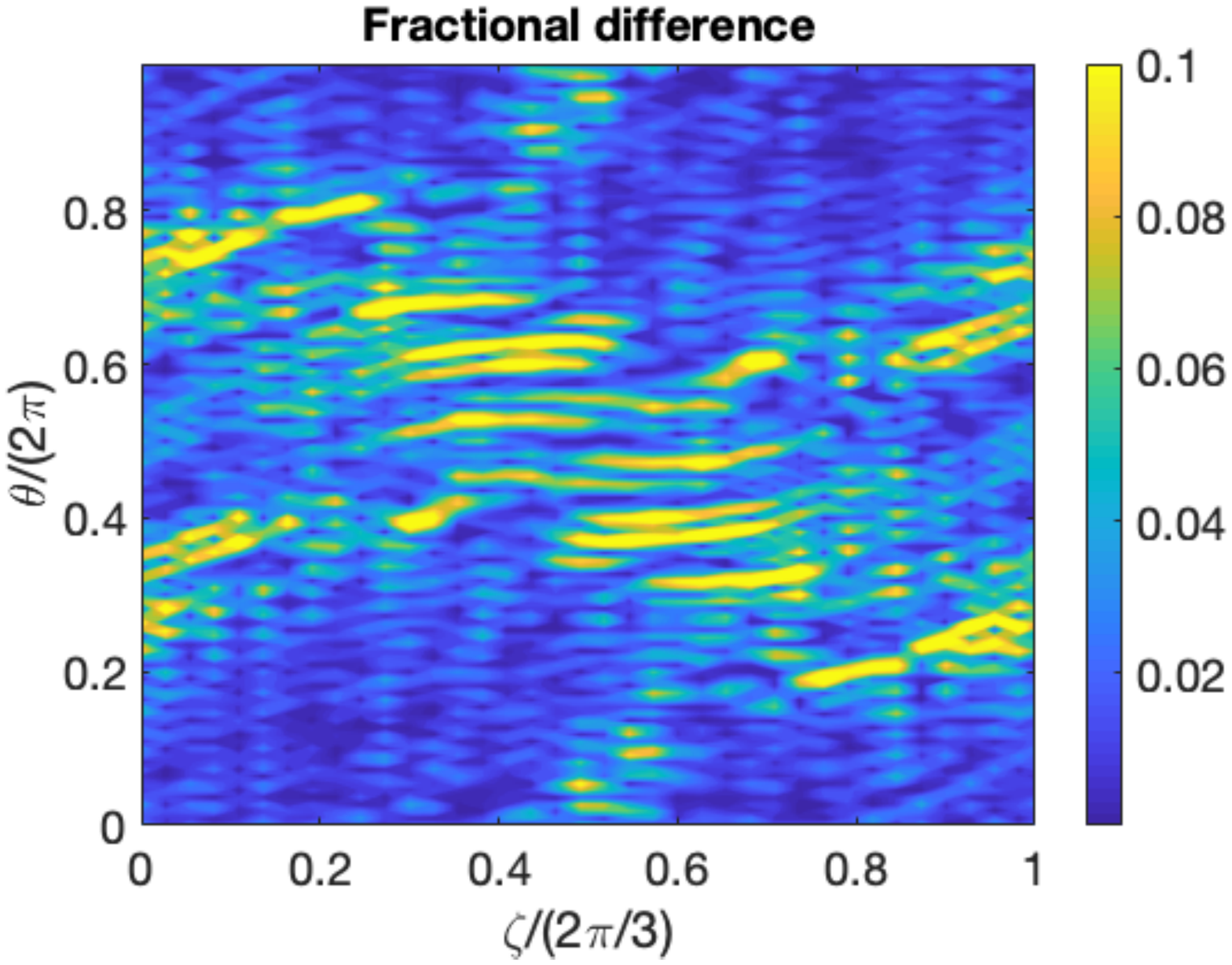}
    \caption{}
    \label{fig:iota_residual}
    \end{subfigure}
    \caption{(a) The shape gradient for $f_{\iota}$ \cref{eq:iota} computed using the adjoint solution \cref{eq:iota_adjoint} (left) and using parameter derivatives (right). (b) The shape gradient computed with the adjoint solution in the $\zeta - \theta$ plane. 
    (c) The fractional difference \eqref{eq:residual} between the shape gradient obtained with the adjoint solution and with parameter derivatives. Again, the results are essentially indistinguishable, as expected.
    } 
    \label{fig:iota_shape_gradient}
\end{figure}

To demonstrate, we use the NCSX stellarator LI383 equilibrium. The toroidal current profile was perturbed with $I_{\Delta} = 715$ A, $\psi_m = 0.1 \psi_0$, and $\psi_w = 0.05 \psi_0$. The shape gradient obtained with the adjoint solution and with the direct approach are shown in \cref{fig:iota_torus}. 
For the direct approach, the shape gradient is computed from parameter derivatives ($\partial f_{\iota}/\partial R_{mn}^c$ and $\partial f_{\iota}/\partial Z_{mn}^s$) using a 8-point stencil with $m \le 18$ and $|n| \le 12$. The fractional difference, $S_{\text{residual}}$, between the two approaches is shown in \cref{fig:iota_residual}, with a surface-averaged value of $2.7\times10^{-2}$.

   The direct approach used 7401 calls to VMEC, while the adjoint only required two. Again, it is apparent that the adjoint method allows the same derivative information to be computed at much lower computational cost. 
   
We find that over much of the surface, the shape gradient is close to zero. A region of large negative shape gradient occurs in the concave region of the plasma surface with adjacent regions of large positive shape gradient. This indicates that ``pinching'' the surface in this region, making it more concave, would increase $\iota$ near the axis. The relationship between shaping and rotational transform is generally quite complex. Further analysis is needed to interpret the shape gradient for $f_{\iota}$.

\subsection{Coil shape gradient for rotational transform}
\label{sec:coil_iota}

The shape gradient of $f_{\iota}$ can also be computed with a free boundary approach. We can again choose a toroidal current profile $\delta I_{T,2} = I_{\Delta} w(\psi)$ for the adjoint problem, where $w(\psi)$ is given by \eqref{eq:weighting}. Using \cref{eq:iota_deriv} and \cref{eq:free_boundary} and noting that $\delta I_{T,2}$ vanishes at the boundary, we find
\begin{gather}
    \delta f_{\iota}(C;\delta \textbf{r}_C) = \frac{1}{2\pi I_{\Delta}}\int_{V_V} d^3 x \, \delta \textbf{J}_{C_1} \cdot \delta \textbf{A}_{V_2}.
\end{gather}
Using \cref{eq:coil_perturb}, this can be written in terms of changes in the positions of coils in the vacuum region,
\begin{gather}
    \delta f_{\iota}(C;\delta \textbf{r}_C) = \frac{1}{2\pi I_{\Delta}} \sum_k \left(  I_{C_k} \int_{C_k} dl \, \delta \textbf{r}_{C_k} (\textbf{x}) \cdot \textbf{t} \times \delta \textbf{B}_2 \right).
    \label{eq:iota_derivative_adjoint}
\end{gather}
When computing the coil shape gradient, the current in each coil is fixed. In arriving at \cref{eq:iota_derivative_adjoint}, we assume that $\delta I_{C_{1,k}} = 0$.  The coil shape gradient is thus
\begin{gather}
    \textbf{S}_k = \frac{I_{C_k} \textbf{t} \times \delta \textbf{B}_2}{2\pi I_{\Delta}}.
    \label{eq:iota_coil_adjoint}
\end{gather}
As anticipated, $\textbf{S}_k$ has no component in the direction tangent to the coil. Evaluating the shape gradient requires computing the adjoint magnetic field at the unperturbed coil locations in the vacuum region. This can be performed with the DIAGNO code \citep{Gardner1990,Lazerson2012}, which employs the virtual casing principle to efficiently compute the fields in the vacuum region due to the plasma current.

\begin{figure}
    \centering
    \includegraphics[trim={3cm 10cm 4cm 10cm},clip,width=0.8\textwidth]{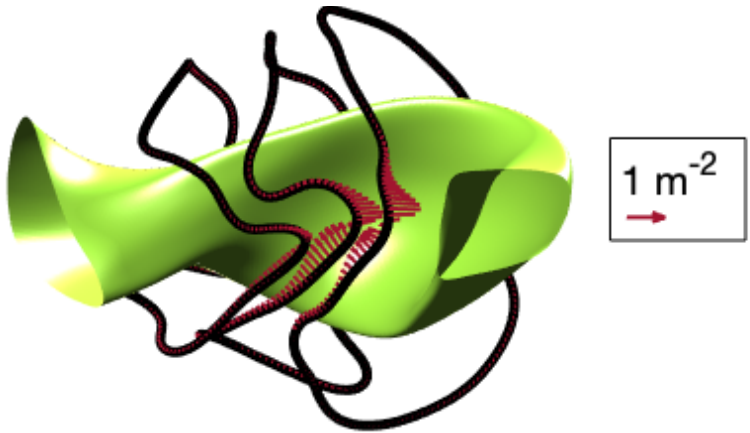}
    \caption{The coil shape gradient for $f_{\iota}$ \cref{eq:iota} computed using the adjoint solution \cref{eq:iota_coil_adjoint} for each of the 3 unique coil shapes (black). The arrows indicate the direction of $\textbf{S}_k$, and their length indicates the local magnitude relative to the reference arrow shown. The arrows are not visible on this scale on the outboard side.
    }
    \label{fig:iota_coil_analytic}
\end{figure}

\begin{figure}
    \centering
    \begin{subfigure}[b]{0.8\textwidth}
    \centering
    \includegraphics[trim={1cm 7cm 1cm 7cm},clip,width=1.0\textwidth]{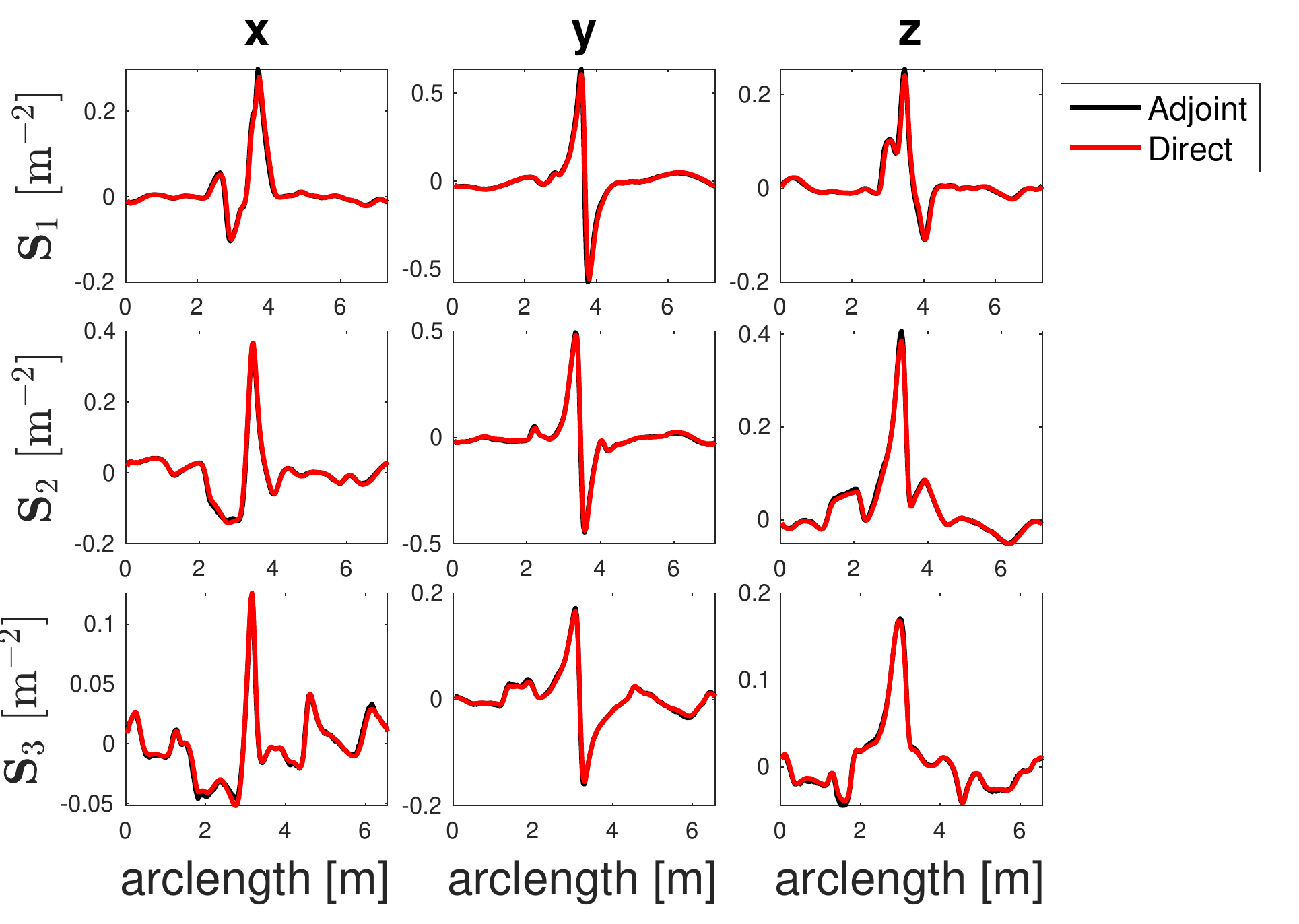}
    \caption{}
    \label{fig:iota_coil_comparison}
    \end{subfigure}
    \begin{subfigure}[b]{0.8\textwidth}
    \includegraphics[trim={1cm 7cm 1cm 7cm},clip,width=1.0\textwidth]{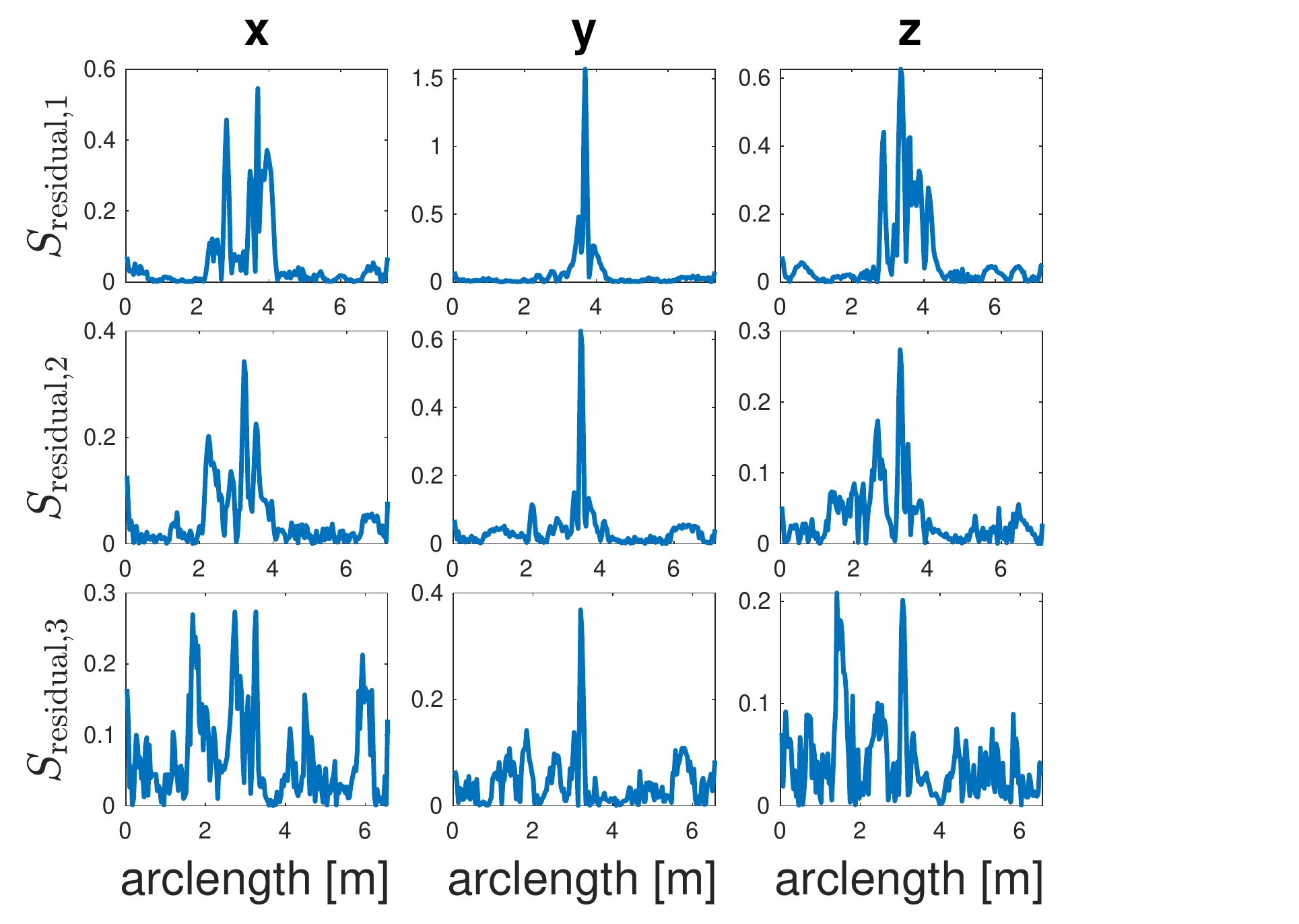}
    \caption{}
    \label{fig:iota_coil_residual}
    \end{subfigure}
    \caption{(a) The Cartesian components of the coil shape gradient for each of the 3 unique modular NCSX coils computed with the adjoint and direct approaches. 
    (b) The fractional difference \eqref{eq:residual_coil} between the shape gradient computed with the adjoint approach and the direct approach is plotted for each Cartesian component and each of the 3 unique coils.}
    \label{fig:iota_residual_coil}
\end{figure}

To demonstrate, we use the NCSX stellarator LI383 equilibrium. The toroidal current profile was perturbed with $I_{\Delta} = 5.7\times 10^5$ A, $\psi_m = 0.1 \psi_0$, and $\psi_w = 0.05 \psi_0$. The shape gradient is computed for each of the three unique modular coils per half period \citep{Williamson2005}, keeping the planar coils fixed. The result obtained with the adjoint solution, $\textbf{S}_{\text{adjoint},k}$, is shown in \cref{fig:iota_coil_analytic}. The shape gradient is also computed with the direct approach, $\textbf{S}_{\text{direct},k}$. For the direct approach, the Cartesian components of each coil are Fourier discretized ($X_{m}$, $Y_{m}$, $Z_{m})$. The numerical derivative with respect to these parameters are computed for $m \leq 45$ using an 8-point stencil. In \cref{fig:iota_coil_comparison} the Cartesian components of the shape gradient computed with the adjoint approach, $S_{\text{adjoint},k}^l$, and with the direct approach, $S_{\text{direct},k}^l$, are shown for each coil. Here $l\in \{x,y,z\}$. The arrows indicate the direction and magnitude of $\textbf{S}_k$ such that if a coil were deformed in the direction of $\textbf{S}_k$, $f_{\iota}$ would increase according to \eqref{eq:shape_gradient_coil}. The direct approach used 6553 calls to VMEC, while the adjoint only required two. In \cref{fig:iota_coil_residual} the fractional difference between the results obtained with the two methods,
\begin{gather}
    S_{\text{residual},k}^l = \frac{|S_{\text{adjoint},k}^l - S_{\text{direct},k}^l|}{\sqrt{\int_{C_k} dl \, \left(S_{\text{adjoint},k}^l\right)^2/\int_{C_k} dl}},
    \label{eq:residual_coil}
\end{gather}
is plotted. The line-averaged values of $S_{\text{residual}}^l$ are $6.1\times 10^{-2}$ for coil 1, $3.8 \times 10^{-2}$ for coil 2, and $4.8 \times 10^{-2}$ for coil 3.

From \cref{fig:iota_coil_analytic} we see that the sensitivity of $f_{\iota}$ to coil displacements is much higher in regions where the coils are close to the plasma surface. The shape gradient points toward the plasma surface in the concave region of the plasma surface, while on the outboard side the sensitivity is significantly lower, again indicating the ``pinching" effect seen in figure \ref{fig:iota_shape_gradient}.

\section{Conclusions}

We have obtained a relationship between 3D perturbations of MHD equilibria that is a consequence of the self-adjoint property of the MHD force operator. The relation allows for the efficient computation of shape gradients for either the outer plasma surface using the fixed boundary adjoint relation  \eqref{eq:fixed_boundary} or for coil shapes using the free boundary adjoint relation \eqref{eq:free_boundary}. The computation of the shape gradient of several stellarator figures of merit has been demonstrated with both the adjoint and direct approach. The application of the adjoint relation provides an $\mathcal{O}(N)$ reduction in CPU hours required in comparison with the direct method of computing the shape gradient, where $N$ is the number of parameters used to describe the shape of the outer boundary or the coils. For fully 3D geometry, $N$ can be $10^2-10^3$. Thus, the application of adjoint methods can significantly reduce the cost of computing the shape gradient for gradient-based optimization or local sensitivity analysis. 

To compute the adjoint equilibria in this work, the full non-linear MHD equilibrium equations are solved using VMEC with the addition of a small perturbation to the current profile, characterized by $I_{\Delta}$, or a small perturbation to the pressure profile, characterized by $\Delta$. These parameters must be tuned carefully such that the perturbation is large enough that the result is not dominated by round-off error, but small enough that non-linear effects do not become important. If a perturbed equilibrium code were instead used, these difficulties could be avoided. However, it is convenient to use the same code for both the unperturbed and adjoint equilibrium.

It should be noted that the adjoint approach we have outlined can not yield an exact analytic shape gradient. Throughout we have assumed the existence of magnetic surfaces as the 3D equilibrium is perturbed. Therefore a code such as VMEC, which minimizes an energy subject to the constraint that surfaces exist, is suitable. Generally VMEC solutions do not satisfy \eqref{eq:force_balance} exactly,  \citep{Nuhrenberg2009}, as they do not account for the formation of islands or current singularities associated with rational surfaces. Furthermore, the the parameters $\Delta$ and $I_{\Delta}$ introduce additional numerical noise. We have demonstrated that the typical difference between the shape gradient obtained with the adjoint method and that computed directly from numerical derivatives is $\lesssim 5\%$. These errors should not be significant for applying the shape gradient to an analysis of engineering tolerances. The discrepancy between the true shape gradient and that obtained numerically, with the adjoint approach or with finite difference derivatives, may become problematic as one nears a local minimum during gradient-based optimization, as the resulting shape gradient may not provide a true descent direction.


For the figure of merit considered in \cref{sec:surf_Beta}, the additional force ($\delta \textbf{F}_2$ in \eqref{eq:perturbed_force_balance}) applied to the adjoint problem can be expressed as a gradient of a scalar pressure; thus an existing equilibrium code could be utilized without modification. This approach could be applied to other quantities of interest. For example, consider a figure of merit which quantifies the vacuum magnetic well,
\begin{gather}
    f_W = \int_{V_P} d^3 x \, w(\psi) V''(\psi),
\end{gather}
where $w(\psi)$ is a weighting function and $V''(\psi) = \partder{}{\psi} \left( \int_0^{2\pi} d \theta \int_0^{2\pi} d \zeta \, \sqrt{g}\right)$. The presence of a magnetic well ($V''(\psi)<0$) has a stabilizing effect on MHD modes \citep{Helander2014,Greene1997} and has been considered in stellarator design \citep{Hirshman1999,Drevlak2018}. Computing the shape gradient of $f_W$ requires solving an adjoint force balance equation with a perturbation to the scalar pressure, similar to the calculation in \cref{sec:surf_Beta}.

For many interesting figures of merit the spatial dependence of the required force is more complicated. Thus, an equilibrium code that allows for an arbitrary force perturbation is needed. One possibility is a generalization of the code ANIMEC \citep{cooper2009} that currently treats anisotropic pressure tensors in the form of a bi-Maxwellian distribution. For example, the shape gradient of the neoclassical particle flux in the $1/\nu$ regime \citep{Nemov1999} can be computed with the addition of a bulk force that takes the form $\delta \textbf{F}_2 = -\nabla \cdot \mathbb{P}_2$ to the adjoint force balance equation. Here $\mathbb{P}_2$ is a pressure tensor that has arbitrary spatial dependence. For several figures of merit, the required additional bulk force does not take the form of the divergence of a pressure tensor. Consider the following figure of merit, which quantifies the departure from quasi-symmetry,
\begin{gather}
    f_{QS} = \int_{V_P} d^3 x \, \left[ \bm{B} \times \nabla \psi \cdot \nabla B - F(\psi) \bm{B} \cdot \nabla B \right]^2 w(\psi),
\end{gather}
where $w(\psi)$ is a weighting function and 
\begin{gather}
    F(\psi) = \frac{(M/N)G(\psi) +I(\psi)}{M/N \iota(\psi) - 1}.
\end{gather}
Here $M$ and $N$ are the mode numbers of the desired quasi-symmetry such that if $B(\psi, M\theta-N\zeta)$ for Boozer angles $\theta$ and $\zeta$, then $f_{QS} = 0$ \citep{Helander2014}. The Boozer covariant components are $G(\psi) =2 I_P/c$ and $I(\psi) = 2I_T/c$, where $I_P$ is the poloidal current outside the $\psi$ surface. Computing the shape gradient of $f_{QS}$ requires the addition of a bulk force to the adjoint force balance equation which does not take the form of the divergence of a tensor pressure. 
These calculations will be reported in a separate publication. 
We anticipate there will be numerous additional applications of this technique for efficient optimization of MHD equilibria. 

\section*{Acknowledgements}
This work was supported by the US Department of Energy through grants DE-FG02-93ER-54197 and DE-FC02-08ER-54964. The computations presented in this paper have used resources at the National Energy Research Scientific Computing Center (NERSC). 

\appendix

\section{Derivation of adjoint relation} 
\label{app:adjoint_relation}
The quantity $U_P = U_{P_1} + U_{P_2}$ consists of two terms, accounting for changes to the vector potential due to MHD perturbations 
\begin{align}
U_{P_1} = \frac{1}{c} \int_{V_P} d^3 x \, \left(\delta\textbf{J}_{1} \cdot \bm{\xi}_2 \times \textbf{B} - \delta \textbf{J}_2\cdot \bm{\xi}_1 \times \textbf{B} \right),
\end{align}
and changes to the rotational transform,
\begin{align}
    U_{P_2} = \frac{1}{c} \int_{V_P} d^3 x \, \left(\delta \Phi_1 \delta \textbf{J}_2 \cdot \nabla \zeta - \delta \Phi_2 \delta \textbf{J}_1 \cdot \nabla \zeta \right). 
    \label{eq:Up2}
\end{align}
The quantity $U_{P_1}$ can be expressed by using \eqref{eq:perturbed_force_balance} and applying the divergence theorem to the pressure gradient terms, 
\begin{multline}
    U_{P_1} = \int_{V_P} d^3 x \, \bm{\xi}_2 \cdot \left( \frac{\textbf{J} \times \delta \textbf{B}_1}{c} + \nabla p\left(\nabla \cdot \bm{\xi}_1\right) + \delta \textbf{F}_1 \right) \\
    - \int_{V_P} d^3 x \, \bm{\xi}_1 \cdot \left( \frac{\textbf{J} \times \delta \textbf{B}_2}{c} +\nabla p \left( \nabla \cdot \bm{\xi}_2 \right) + \delta \textbf{F}_2 \right).
    \label{eq:Up1}
\end{multline}
We will define $\delta \hat{\textbf{B}}_{1,2} = \nabla \times \left(\bm{\xi}_{1,2} \times \textbf{B} \right)$ such that $\delta \textbf{B}_{1,2} = \delta \hat{\textbf{B}}_{1,2} - \nabla \delta \Phi_{1,2} \times \nabla \zeta$.
The terms in \eqref{eq:Up1} due to $\delta \hat{\textbf{B}}_{1,2}$ can be evaluated using $\textbf{J} = J_{||} \textbf{b} + c\textbf{b}\times\nabla p/B$ and \eqref{eq:force_balance},
\begin{multline}
    \frac{1}{c}\int_{V_P} d^3 x \, \left( \bm{\xi}_2 \cdot \textbf{J} \times \delta \hat{\textbf{B}}_1 - \bm{\xi}_1 \cdot \textbf{J} \times \delta \hat{\textbf{B}}_2\right) = \int_{V_P} d^3 x \, \frac{J_{||}}{cB}\nabla \cdot \left( \left(\bm{\xi}_1 \times \textbf{B} \right) \times \left( \bm{\xi}_2 \times \textbf{B} \right) \right)\\ + \int_{V_P} d^3 x \, \frac{1}{B} \left( \left(\bm{\xi}_2 \cdot \nabla p \right) \textbf{b} \cdot \delta \hat{\textbf{B}}_1  - \left(\bm{\xi}_1 \cdot \nabla p \right) \textbf{b} \cdot \delta \hat{\textbf{B}}_2 \right).
    \label{eq:Upp}
\end{multline}
The first term in \eqref{eq:Upp} can be simplified using $\nabla \cdot \textbf{J} = 0$ and noting that the perturbation can be written as $\bm{\xi}_{1,2} = \xi^{\psi}_{1,2} \nabla \psi + \xi^{\perp}_{1,2} \textbf{b} \times \nabla \psi$.
Applying the identity $\textbf{b} \cdot \delta \hat{\textbf{B}}_{1,2} = - B^2 \nabla \cdot \bm{\xi}_{1,2} - \bm{\xi}_{1,2} \cdot \nabla B^2 - 4 \pi \bm{\xi}_{1,2} \cdot \nabla p$ to the second term, the following expression can be obtained, 
\begin{multline}
    \frac{1}{c}\int_{V_P} d^3 x \, \left( \bm{\xi}_2 \cdot \textbf{J} \times \delta \hat{\textbf{B}}_1 - \bm{\xi}_1 \cdot \textbf{J} \times \delta \hat{\textbf{B}}_2\right) = \\ \int_{V_P} d^3 x \, \left(\left(\nabla \cdot \bm{\xi}_2\right) \bm{\xi}_1 \cdot \nabla p - \left( \nabla \cdot \bm{\xi}_1 \right) \bm{\xi}_2 \cdot \nabla p \right).
\end{multline}
Hence we obtain the following expression for $U_{P_1}$,
\begin{gather}
    U_{P_1} = \int_{V_P} d^3 x \, \left( \bm{\xi}_2 \cdot \delta \textbf{F}_1 - \bm{\xi}_1 \cdot \delta \textbf{F}_2\right) - \frac{1}{c}\int_{V_P} d^3 x \, \left( \delta \iota_1 \bm{\xi}_2 \cdot \nabla \psi  - \delta \iota_2 \bm{\xi}_1 \cdot \nabla \psi  \right) \textbf{J} \cdot \nabla \zeta.
\end{gather}
We now consider $U_{P_2}$ defined in \eqref{eq:Up2}.
Applying \eqref{eq:delta_I} for the change in toroidal current, integrating by parts in $\psi$, and combining the expressions for $U_{P_1}$ \eqref{eq:Up1} and $U_{P_2}$ \eqref{eq:Up2}, we obtain
\begin{multline}
    U_P = \int_{V_P} d^3 x \, \left(\bm{\xi}_2 \cdot \delta \textbf{F}_1 - \bm{\xi}_1 \cdot \delta \textbf{F}_2\right) + \frac{2\pi}{c} \int_{V_P} d \psi \, \left( \delta \Phi_1 \der{\delta I_{T,2}}{\psi} - \delta \Phi_2 \der{\delta I_{T,1}}{\psi} \right) \\
    - \frac{1}{c} \int_{S_P} d^2 x \, \left(\delta \Phi_1 \bm{\xi}_2 - \delta \Phi_2 \bm{\xi}_1 \right) \cdot \textbf{n} \textbf{J} \cdot \nabla \zeta.
    \label{eq:Upb}
\end{multline}
Next we combine $U_P$ \eqref{eq:Upb} with $U_B$ \eqref{eq:Ub} and add $U_c$ \eqref{eq:Uc} to obtain the free boundary adjoint relation \eqref{eq:free_boundary}.

To obtain the fixed-boundary adjoint relation, the integral over the plasma volume \eqref{eq:Upa} can be related to a surface integral by applying the divergence theorem, \cref{eq:Up_bound}. Using \eqref{eq:delta_A1} and applying several vector identities,
\begin{multline}
    U_P = -\frac{1}{4\pi} \int_{S_P} d^2 x \, \textbf{n} \cdot \left( \bm{\xi}_1 \delta \textbf{B}_2 \cdot \textbf{B} - \bm{\xi}_2 \delta \textbf{B}_1 \cdot \textbf{B} \right) \\
    - \frac{1}{4\pi} \int_{S_P} d^2 x \, \left( \delta \Phi_2 \delta \textbf{B}_1 - \delta \Phi_1 \delta \textbf{B}_2 \right) \cdot \nabla \zeta \times \textbf{n}.
    \label{eq:Uc+Ub}
\end{multline}
Using the expression for $U_P$, \eqref{eq:Upb}, expressing the second term in \eqref{eq:Uc+Ub} as a perturbed current using \eqref{eq:delta_I}, the fixed boundary adjoint relation \eqref{eq:fixed_boundary} is obtained.

\section{Alternate derivation of fixed-boundary adjoint relation}
\label{appendix:self-adjointness}

The MHD force operator,
\begin{gather}
    \textbf{F}(\bm{\xi}_{1,2}) = \frac{\textbf{J} \times \left(\nabla \times \left(\bm{\xi}_{1,2} \times \textbf{B} \right)\right)}{c} + \frac{\nabla \times \left( \nabla \times \left(\bm{\xi}_{1,2} \times \textbf{B} \right) \right) \times \textbf{B}}{4\pi} + \nabla \left(\bm{\xi}_{1,2} \cdot \nabla p \right),
\end{gather}
possesses the following self-adjointness property \citep{Bernstein1958, Goedbloed2004},
\begin{gather}
    \int_{V_P} d^3 x \, \left( \bm{\xi}_2 \cdot \textbf{F}_1 - \bm{\xi}_1 \cdot \textbf{F}_2 \right) = \frac{1}{4\pi}\int_{S_P} d^2 x \, \textbf{n} \cdot \left( \bm{\xi}_1 \textbf{B} \cdot \delta \hat{\textbf{B}}_2 - \bm{\xi}_2 \textbf{B} \cdot \delta \hat{\textbf{B}}_1 \right) ,
    \label{eq:self_adjointness}
\end{gather}
where $\delta \hat{\textbf{B}}_{1,2} = \nabla \times \left(\bm{\xi}_{1,2} \times \textbf{B}\right)$ is the perturbed field corresponding to the MHD perturbations. For perturbations described by \cref{eq:delta_A1,eq:delta_A2,eq:delta_I,eq:perturbed_force_balance}, we have the following force operator,
\begin{gather}
    \textbf{F}(\bm{\xi}_{1,2}) = \frac{\textbf{J} \times \left( \nabla \delta \Phi_{1,2} \times \nabla \zeta \right)}{c} + \frac{\nabla \times \left( \nabla \delta \Phi_{1,2} \times \nabla \zeta \right) \times \textbf{B}}{4\pi}  - \delta \textbf{F}_{1,2}.
    \label{eq:force_operator}
\end{gather}
Using \cref{eq:force_operator} and several vector identities, the left hand side of \cref{eq:self_adjointness} can be written as
\begin{multline}
    \int_{V_P}d^3 x \, \left( \bm{\xi}_2 \cdot \textbf{F}_1 - \bm{\xi}_1 \cdot \textbf{F}_2 \right) = \frac{1}{c}\int_{V_P} d^3 x \, \left( \delta \iota_1 \bm{\xi}_2 - \delta \iota_{2} \bm{\xi}_1 \right)\cdot \nabla \psi \textbf{J} \cdot \nabla \zeta \\ - \frac{1}{4\pi} \int_{V_P} d^3 x \, \nabla \psi \times \nabla \zeta \cdot \left( \delta \iota_1 \delta \hat{\bm{B}}_2 - \delta \iota_2 \delta \hat{\bm{B}}_1 \right)\\
    - \frac{1}{4\pi} \int_{S_P} d^2 x \, \left(\bm{\xi}_2 \delta \iota_{1} - \bm{\xi}_1 \delta \iota_2 \right) \cdot \textbf{n} \left( \nabla \psi \times \nabla \zeta \cdot \textbf{B}\right) - \int_{V_P} d^3 x \, \left( \bm{\xi}_2 \cdot \delta \textbf{F}_1 - \bm{\xi}_1 \cdot \delta \textbf{F}_2 \right).
    \label{eq:self_adjoint_1}
\end{multline}
In arriving at \eqref{eq:self_adjoint_1}, we use $\textbf{J} \cdot \nabla \psi = 0$.
Using \cref{eq:delta_I} to re-express the first two terms on the right hand side,
\begin{multline}
  \int_{V_P}d^3 x \, \left( \bm{\xi}_2 \cdot \textbf{F}_1 - \bm{\xi}_1 \cdot \textbf{F}_2 \right) =  \frac{2 \pi}{c} \int_{V_P} d \psi \, \left( \delta I_{T,2} \delta \iota_{1} - \delta I_{T,1} \delta \iota_{2} \right) \\ - \frac{1}{4\pi} \int_{S_P} d^2 x \, \left( \bm{\xi}_2 \delta \iota_{1}  - \bm{\xi}_1 \delta \iota_{2} \right) \cdot \textbf{n} \left( \nabla \psi \times \nabla \zeta \cdot \textbf{B} \right)
  - \int_{V_P} d^3 x \, \left( \bm{\xi}_2 \cdot \delta \textbf{F}_1 - \bm{\xi}_1 \cdot \delta \textbf{F}_2 \right).
\end{multline}
Using \cref{eq:delta_A1,eq:self_adjointness} we obtain \cref{eq:fixed_boundary}. 

\section{Interpretation of the displacement vector}
\label{app:displacement}
For MHD perturbations such that $\delta \textbf{B} = \nabla \times \left( \bm{\xi} \times \textbf{B} \right)$ the displacement can be interpreted as a vector describing the motion of a field lines. Thus a normal perturbation to the surface of the plasma as in \eqref{eq:shape_gradient_surface} can be expressed in terms of the displacement vector,
\begin{gather}
    \delta f(S_P;\bm{\xi}) = \int_{S_P} d^2 x \, S \bm{\xi} \cdot \textbf{n}. 
    \label{eq:shape_gradient_surface_xi}
\end{gather}
For perturbations that allow for changes in the rotational transform it remains to be shown that a similar relation can be found. 

As we require that $\psi$ remain a flux surface label in the perturbed equilibrium, the Lagrangian perturbation to $\psi$ at fixed position is 
\begin{gather}
    \delta \psi = - \delta \textbf{r} \cdot \nabla \psi. 
\end{gather}
The perturbed magnetic field, $\textbf{B}' = \textbf{B} + \delta \textbf{B}$ must remain tangent to $\psi' = \psi + \delta \psi$ surfaces; thus to first order in the perturbation,
\begin{gather}
    0 = \textbf{B}' \cdot \nabla \psi' = \textbf{B} \cdot \nabla \delta \psi + \delta \textbf{B} \cdot \nabla \psi. 
\end{gather}
Applying the form for the perturbed field allowing for changes in the rotational transform, $\delta \textbf{B} = \nabla \times \left( \bm{\xi} \times \textbf{B} - \delta \Phi \nabla \zeta \right)$, and using several vector identities, the following condition is obtained
\begin{gather}
    \textbf{B} \cdot \nabla \left( \delta \textbf{r} \cdot \nabla \psi \right) = \textbf{B} \cdot \nabla \left( \bm{\xi} \cdot \nabla \psi \right).
\end{gather}
This implies that $\delta \textbf{r} \cdot \nabla \psi = \bm{\xi} \cdot \nabla \psi + F(\psi)$, where $F(\psi)$ is some flux function which can be determined by requiring that the perturbation to the toroidal flux as a function of $\psi$ vanishes, $\delta \Phi_T(\psi) = 0$. 

The perturbed toroidal flux through a surface labeled by $\psi$ contains two terms, corresponding to the flux of the unperturbed field through the perturbed surface and the perturbed field through the unperturbed surface, 
\begin{gather}
    \delta \Phi_T(\psi) = \int_{\partial S_T(\psi)} d \theta \, \sqrt{g}  \delta \textbf{r} \cdot \nabla \psi \textbf{B} \cdot \nabla \zeta + \int_{S_T(\psi)} d \psi d \theta \, \sqrt{g} \delta \textbf{B} \cdot \nabla \zeta. 
\end{gather}
Using the form for $\delta \textbf{B}$,  applying the divergence theorem, and noting that $\textbf{B} \cdot \nabla \zeta = \sqrt{g}^{-1}$, the following condition is obtained,
\begin{gather}
    \delta \Phi_T(\psi) = \int_0^{2\pi} d \theta \, \left(\delta \textbf{r} \cdot \nabla \psi - \bm{\xi} \cdot \nabla \psi \right). 
\end{gather}
By requiring that $\delta \Phi_T(\psi) = 0$, we find that $F(\psi) = 0$. Thus we can express shape gradients in the form of \eqref{eq:shape_gradient_surface_xi} even when the rotational transform is allowed to vary. 

\bibliographystyle{jpp}
\bibliography{bibliography}
\end{document}